\documentclass[aps,pra,twocolumn,superscriptaddress,floatfix]{revtex4-2}
\usepackage{amsmath}
\usepackage{amssymb}
\usepackage{graphicx}
\usepackage{color}
\usepackage{bm}
\usepackage[colorlinks=true,citecolor=blue]{hyperref}

\newcommand{\ket}[1]{{\vert #1 \rangle}}
\newcommand{\ketbra}[2]{{\vert #1 \rangle\langle #2\vert}}

\newcommand{\braket}[2]{\langle{#1}|{#2}\rangle}
\newcommand* {\bra}[1]{\ensuremath{\langle {#1} |}}

\begin{document}
	\title{
	Entangling operations in nonlinear two-atom Tavis-Cummings models
	}

	\author{Roc\'io G\'omez-Rosas} 
	\affiliation{Instituto de F\'isica, Benem\'erita Universidad Aut\'onoma de Puebla, Apartado Postal J-48, Puebla 72570, Mexico}
	\author{Carlos A. Gonz\'alez-Guti\'errez}
	\affiliation{Instituto  de  Nanociencia  y  Materiales  de  Arag\'on  (INMA) and Departamento de F\'isica de la Materia Condensada, CSIC-Universidad  de  Zaragoza,  Zaragoza 50009, Spain}
	\author{Juan Mauricio Torres}
	\email{jmtorres@ifuap.buap.mx}
	\affiliation{Instituto de F\'isica, Benem\'erita Universidad Aut\'onoma de Puebla, Apartado Postal J-48, Puebla 72570, Mexico}

	\date{\today}
	\begin{abstract}
		We derive an analytical approximate solution of the 
		time-dependent state vector in terms of material Bell states
		and coherent states of the field for a generalized two-atom Tavis-Cummings model with nonlinear intensity dependent matter-field interaction. Using this solution, we obtain simple expressions
		for the atomic concurrence and purity in order to study the 
		entanglement in the system at specific interaction times. 
		We show how to implement entangling atomic operations through 
		measurement of the field. We illustrate how these operations can lead to a complete Bell measurement. Furthermore,
		when considering two orthogonal states of the field as levels
		of a third qubit, it is possible to implement a unitary three-qubit gate capable of generating authentic tripartite entangled
		states such as the Greenberger–Horne–Zeilinger (GHZ) state and the $W$-state. As an example of the generic model, we present an ion-trap setting employing the
		quantized mode of the center of mass motion instead the photonic field, showing that the implementation of realistic entangling operations from intrinsic nonlinear matter-field interactions is indeed possible.
	\end{abstract}
	\maketitle

\section{Introduction}

 Entangling quantum gates are crucial in quantum information and quantum computation protocols such
as quantum teleportation, superdense coding, and Shor's algorithm \cite{Nielsen00,Cirac2001}.  
For the implementation of these gates in atomic qubits, 
cavity quantum electrodynamics (QED) has played an important role, as generating and controlling entangled states has become an  experimental reality \cite{Zheng2000a,Raimond2001,Osnaghi2001,Ritter2012,Noelleke2013,Casabone2013,Reiserer2015,Walther2006}.  
Some of these concepts and results have been shared with other 
settings  \cite{Leibfried2003a,Pedernales2015,Blais2021}.
The celebrated Cirac-Zoller controlled-NOT (CNOT) gate
 is an example in the context of ion traps, where a Jaynes-Cummings interaction \cite{Jaynes1963} between electronic levels of the ions and its mechanical oscillatory mode has been exploited in order to mediate the interaction
 between the ions \cite{Cirac1995,Schmidt-Kaler2003}. 
 Similar applications have been found in the context of superconducting systems, where artificial
 atoms can be tailored to specific needs \cite{Majer2007,Fink2009,Linke2017}.
 While certain problems seem to be solved,  it is important to offer other advantageous alternatives for different experimental settings. 
 For instance,
 the M\o{}lmer-S\o{}rensen entangling gate in ion traps does not
 require  ground state cooling as the Cirac-Zoller gate \cite{Molmer1999,Haeffner2008,Roos2008}.

Recent proposals offer
new perspectives exploiting  the multiphoton
 regime in cavity-QED such as the hybrid quantum repeater utilizing dispersive and resonant interactions of matter qubits and coherent light states \cite{VanLoock2006,Ladd2006,Bernad2013,Bernad2017}. 
 It has been shown that using the Jaynes-Cummings interaction assisted with multiphoton states, it is possible to implement a nonunitary entangling 
 operation replacing the CNOT gate in an entanglement purification protocol \cite{Bernad2016a,Torres2016a}. 
 Furthermore, exploiting a two-photon interaction with multiphoton states, 
  it is in principle possible to implement a complete Bell measurement (BM) by measuring the state of the field \cite{Gonzalez-Gutierrez2018a}. 
 An important feature to achieve this BM is that the two-photon interaction model presents
 perfect revivals of Rabi oscillations in the system observables, in contrast to the  Tavis-Cummings interaction \cite{Tavis1968b}, where these revivals broaden in time 
 \cite{Torres2014,Torres2016b}.
 A natural question is whether other models with perfect revivals of Rabi oscillations could also be useful in this type of protocols. 
 This is relevant in an ion-trap implementation of the model, as single phonon processes are simpler to achieve than two-phonon ones \cite{Leibfried2003a}. Furthermore,
 large coherent states in the motional degree of trapped ions are nowadays accessible and controllable \cite{Johnson2017,McDonnell2007,Alonso2016}, making it an interesting candidate to implement
 multiphoton regime machinery from cavity QED as multiphonon ion-trap protocols. 
 
 In this paper, we study a generalized version of the two-atom Tavis-Cummings model with a nonlinear 
 matter-field interaction. 
 We derive an approximate analytical solution of
 the time-dependent state vector given in terms of material Bell states
 and coherent states of the field. We find  conditions where this approximation remains valid and where perfect
 revivals of Rabi oscillations occur. 
 The simple and general
 form of our solution allows us to study the entanglement in the 
 system and to 
 generate entangling two-qubit and three qutbit quantum operations. 
 We present a viable realization  in an ion-trap setup, where the photonic oscillator is replaced by the
 center of mass motion of the ions.

The manuscript is organized as follows. 
In Sec.~\ref{Model} we introduce a generic nonlinear two-atom Tavis-Cummings model. We identify constants of motion, show its full solvability, and derive a compact
analytical approximate solution.
In Sec.~\ref{Specific} we present three examples of the generic model and we propose the implementation in 
an ion-trap setting.
In Sec. \ref{Dynamical} we study the dynamical features and numerically test our approximate solution. 
We study the entanglement in the system in Sec. \ref{Entanglement}, where we find approximate analytical
expressions for concurrence and purity of the atomic state. Based on our approximate solution, 
in Sec. \ref{QuantumOperations}, we present the implementation of entangling operations for the two and three-qubit case, together with a Bell measurement protocol using a second quantized mode.

\section{Generalized Two-atom Tavis-Cummings model}
\label{Model}
In this section we present the Hamiltonian of a generalized version of the Tavis-Cummings model \cite{Tavis1968b} with a nonlinear
intensity dependent coupling. 
We identify constants of motion that lead to an exact solvability. 
Similar general models  have already been 
considered in the past and their exact solution is known \cite{Buck1981,Kochetov1987,Chaichian1990,Bonatsos1993,Rybin1999,SantosSanchez2016,Vogel1995,Maldonado2020}. 
However, here we are interested
in presenting a general approximate solution for initial coherent states with large mean number of quanta
that is especially convenient to analyze the entanglement in the system as it is expressed
in terms of material Bell states and coherent states of the field.

\subsection{Hamiltonian and exact solvability}
We consider the following Hamiltonian describing two two-level atoms  resonantly interacting
with a quantized harmonic oscillator
\begin{equation}
	\label{eq:Hamiltonian}
    H=\hbar\omega I +V,\quad 
    I=a^\dagger a+S_z.
\end{equation}
The free Hamiltonian has been expressed in terms of the operator $I$ that 
represents the number of excitations in the system. In the present case $I$ commutes with the 
intensity dependent interaction operator
\begin{equation}
	\label{eq:V}
 V\equiv V_a=\hbar \Omega
\left(
f(a^\dagger a)aS_++a^\dagger f(a^\dagger a) S_-
\right).
\end{equation}
We have employed the creation and annihilation
operators of the oscillator, $a$ and $a^\dagger$. In Sec. \ref{Specific} we will specify the nature of the oscillator, that will be considered optical or mechanical
for different particular models.
The interaction includes the intensity
dependent function $f(a^\dagger a)$ leading 
to a nonlinear atom-field interaction. 
We have also used the notation $V_a$ in order to stress the dependence on the specific
mode operator $a$, as it will  probe useful when we introduce a second mode and its operators $b$ and $b^\dagger$ in Sec. \ref{QuantumOperations}.

As for the electronic degrees of freedom of the atoms, we have introduced the following  operators
\begin{align}
	\label{eq:atomoperators}
	S_-&=\ketbra{\rm g}{\rm e}_1+\ketbra{\rm g}{\rm e}_2, \quad S_+=S_-^\dagger
	\nonumber\\
	S_z&=\frac{1}{2}
	\Big(
	\ket{\rm e}\bra{\rm e}_1+\ketbra{\rm e}{\rm e}_2-
	\ketbra{\rm g}{\rm g}_1-\ketbra{\rm g}{\rm g}_2
	\Big),
\end{align}
where
$\ket{\rm e}_1$ ($\ket{\rm e}_2$) and $\ket{\rm g}_1$ ($\ket{\rm g}_2$) are the excited and ground states of
the first (second) atom.
The energy difference between the atomic levels is given by $\hbar\omega$ and coincides
with a single quantum unit of energy of the oscillator. Furthermore, $\hbar\Omega$ represents the
coupling energy between the internal states of the atom and the oscillator degree of freedom. 
The resonance condition implies a second  
constant of motion, namely
$\bm S^2=(S_+S_-+S_-S_+)/2+S_z^2$. If each two-level system
is regarded as a pseudospin, then the operator $\bm S$ plays the role of an
adimensional total
pseudospin operator. The existence of these two constants of motion implies
that the eigenstates of the Hamiltonian have to be simultaneous eigenstates of $I$ and 
$\bm S^2$. Noting this fact, it is natural to work out the problem in the following
basis
\begin{align}
\label{eq:basis}
 \ket{\varphi^{n}}&=\ket{\Psi^-}\ket{n},&
 \ket{\varphi_{-1}^{n}}&=\ket{\rm gg}\ket{n+1},\\
\ket{\varphi_0^{n}}&=\ket{\Psi^+}\ket{n}, & \ket{\varphi_1^{n}}&=\ket{\rm ee}\ket{n-1},
    \nonumber
\end{align}
where we have employed two of the  Bell states
\begin{equation}
    \ket{\Psi^\pm}=\frac{\ket{\rm ge}\pm\ket{\rm eg}}{\sqrt2},\quad
   \ket{\Phi^\pm}=\frac{\ket{\rm gg}\pm\ket{\rm ee}}{\sqrt2}.
\end{equation} 
In the above definitions of the atomic states, we have used the convention of labeling the first atom always to
the left, for instance $\ket{\rm e}_1\ket{\rm g}_2=\ket{\rm eg}$.

The states in Eq. \eqref{eq:basis} are eigenstates of $I$ with eigenvalue $n$
that  takes values from $-1$ to $\infty$. 
These states also fulfill the eigenvalue equations
$\bm S^2\ket{\varphi^n}=0$ and 
$\bm S^2\ket{\varphi_l^n}=1$, where $l\in\{-1,0,1\}$.
This implies that the state  $\ket{\varphi^n}$ is an 
eigenstate of the Hamiltonian, as it is the only one with eigenvalues $n$
for $I$ and $0$ for $\bm S^2$. The remaining three states, for fixed $n$, share
 the same eigenvalue for $\bm S^2$ and therefore form
 a disconnected block of the interaction Hamiltonian $V$. The matrix representation of each block with fixed $n$
 can be expressed as 
\begin{equation}
	\label{eq:BlockV}
    V^{(n)}=\hbar\left(
    \begin{array}{ccc}
       0&\Omega_{n} &0  \\
         \Omega_{n}&   0&\Omega_{n-1}\\
         0&\Omega_{n-1}& 0
         \end{array}
    \right),
\end{equation}
where $\Omega_n$ is a real valued parameter dependent on $n$ and given by the follwowing expression
\begin{equation}
\label{eq:Omega}
\Omega_n= \Omega \sqrt2\bra{n+1}a^\dagger f(a^\dagger a)\ket{n}.
\end{equation}
The nonzero eigenvalues for each subspace can be computed exactly and are simply  given by $E_{n,\pm}=\pm\hbar\sqrt{\Omega_n^2+\Omega_{n-1}^2}$.
The eigenvectors can also be evaluated exactly in closed-form, however, we will resort on approximations
that will probe useful, especially for analyzing the atomic state  when the field is initially prepared in a coherent state
with a large mean excitation value.

The fact that one of the eigenfrequencies in this $3\times 3$ block is zero implies the existence of an additional invariant family of states together with $\ket{\Psi^-}\ket{n}=\ket{\varphi^n}$. This feature is lost in the off-resonant case, where the atomic transition differs from the frequency of the oscillator. In this case,
additional terms appear in the diagonal of  Eq. \eqref{eq:BlockV} leading to
three non-zero eigenfrequencies.

\subsection{Time-dependent state vector}
\label{ssec:statevector}
In order to simplify the calculations, we choose to work 
in an interaction picture with respect to the 
free energy $\hbar \omega I$ that includes another time-independent transformation. In particular, the state vector in this frame is given by 
\begin{equation}
	\label{eq:intpic}
\ket{\Psi(t)}=e^{-iI\phi}e^{iI\omega t}\ket{\Psi(t)}_{\rm S},
\end{equation}
where $\ket{\Psi(t)}_{\rm S}$ is the state vector in the Schrödinger or laboratory frame.
The real parameter  $\phi$ is the phase of the initial state of the field that is assumed to be prepared in the Schrödinger picture in an arbitrary coherent state
\begin{equation}
	\label{eq:cohstate}
	\ket{\alpha e^{i\phi}}=\sum_{n=0}^{\infty}p_{n}e^{i n\phi}\ket{n}, \,\,\,\,
	p_n=e^{-|\alpha|^2/2}
	\frac{\alpha^{n}}{\sqrt{n!}}.
\end{equation}
With the choice of interaction picture as in Eq. \eqref{eq:intpic}, we have encoded the phase of the coherent state
in the unitary operator $e^{-iI\phi}$ that acts in a straightforward way on each subspace of the constant of motion $I$. 
We can restrict our analysis to non-negative values of $\alpha$
without loss of generality.
In this way, we have exploited the commutativity of the constant $I$ with the interaction $V$ in order to simplify the problem. As $\alpha$ is taken real in this work, the mean number of quanta is given by 
$N=\langle{a^{\dagger}a}\rangle=\alpha^2$.

 For the total initial state of the system,  we assume a pure product state of the form 
$\ket{\Psi(0)}=\ket{\psi}\ket\alpha$, where the two atoms are allowed to start in
an arbitrary pure state $\ket\psi$, namely
\begin{align}
	\label{eq:psiin}
	\ket\psi=&
	c_-\ket{\Psi^-}+
	c_+\ket{\Psi^+}+
	d_-\ket{\Phi^-}+
	d_+\ket{\Phi^+}.
\end{align}
We have chosen to write the initial states in terms of Bell states for later convenience. However,
the basis of Eq. \eqref{eq:basis}, in which the Hamiltonain is block-diagonal, contains  two Bell states 
and two bare levels of the atoms. For this reason and in order to keep track 
of the calculations, it is useful to relate the following initial probability amplitudes 
$	d_-\ket{\Phi^-}+
d_+\ket{\Phi^+}=
c_{\rm g}\ket{\rm gg}+
c_{\rm e}\ket{\rm ee}$, where
\begin{equation}
	d_\pm=\frac{c_{\rm g}\pm c_{\rm e}}{\sqrt2}.
\end{equation}
Note that with the transformation in Eq. \eqref{eq:intpic}, the intial atomic state is given in the laboratory frame
as $\ket{\psi}_{\rm S}=e^{iS_z\phi}\ket\psi$.

The solution to the Schrödinger equation in the interaction picture defined in Eq. \eqref{eq:intpic} is given by
$\ket{\Psi(t)}=e^{-iV t/\hbar}\ket{\Psi(0)}$. Using the
basis states in Eq. \eqref{eq:basis} we can formally expand the solution of the time-dependent state vector as
\begin{align}
\label{eq:Psit}
&\ket{\Psi(t)}=
c_-\ket{\Psi^-}\ket{\alpha}+
\sum_{n=-1}^\infty
\sum_{l=-1}^{D_n}
C_{n,l}(t)\ket{\varphi^{n}_l}
\end{align}
with the limit in the second sum $D_n=1-\delta_{n,0}-2\delta_{n,-1}$ given in terms of the Kronecker delta.
This limit depends on the value of $n$ and takes into account
that for $n=-1$ there is only one state in the basis (Eq. \eqref{eq:basis} without $\ket{\varphi^n}$), two states for $n=0$, and three states for $n\ge2$. These states for low $n$
will have no significant contribution in the limit of high number of excitation as $e^{-\alpha^2/2}\simeq 0$. Furthermore,
we have used the fact that $\ket{\varphi^n}$ is eigenstate of $V$ with zero eigenvalue and therefore its probability amplitude 
remains constant as $c_-p_n$.
At $t=0$ one has the initial probability amplitudes
$C_{n,-1}(0)=p_{n+1}c_g$,
$C_{n,0}(0)=p_{n}c_+$,
and $C_{n,1}(0)=p_{n-1}c_e$. As the system is exactly solvable, it is possible to obtain exact analytical expressions for all
the probability amplitudes in Eq. \eqref{eq:Psit} using the exact form of the evolution operator presented in Appendix \ref{Exact}.
In particular, the time-evolution of initial Fock states of the field can be evaluated in a straightforward way using these expressions.

In order to obtain manageable expressions we will resort on three 
approximations. In the first one, we make the replacements $\Omega_{n}\to\Omega_{n-1/2}$ and  $\Omega_{n-1}\to\Omega_{n-1/2}$ in Eq.
\eqref{eq:BlockV}.  In this way, the eigenvectors of $V^{(n)}$ are independent of $n$.
Provided that $|\Omega_n-\Omega_{n-1}|\ll \Omega_n$, the neglected part can be considered as a small perturbation. 
This is indeed the case, for instance, when $\Omega_n\propto n$ or $\Omega_n\propto \sqrt n$. However, we will see later that $\Omega_n$ might have a non monotonic dependence on $n$, but the condition
might be fulfilled for a specific interval outside of which the distribution
$p_n$ in Eq. \eqref{eq:cohstate} presents vanishing small contributions. With the first approximation, one can find that the nonzero eigenenergies are given
by 
\begin{equation}
	\label{eq:omegaeig}
	E_\pm^{(n)}\simeq\pm\hbar\omega_n,\quad 
	\omega_n=\sqrt{2}|\Omega_{n-1/2}|,
\end{equation}
where we have introduced the approximate eigenfrequencies $\omega_n$.
The second approximation is applied to the Poissonian distribution
in the coherent states, namely 
$p_{n-1}\simeq p_n\simeq p_{n+1}$, which relies on the condition of having
a large mean  number of quanta $N\gg1$. Using these two approximations, one
is lead to the following form of the time-dependent  probability amplitudes 
\begin{align}
\label{eq:Cs}
C_{n,l}(t)&\simeq\left[\frac{c_+-d_+}{2(-1)^l}e^{i\omega_n t}+\frac{d_++c_+}{2}
e^{-i\omega_n t}-ld_-\right]\frac{p_{n-l}}{\sqrt2},
\end{align}
with $l\in\{-1,0,1\}$.
The third approximation is made to the eigenfrequencies in 
\eqref{eq:omegaeig} by Taylor expanding around the mean photon
number $N$ as
\begin{equation}
\label{eq:taylor}
	\omega_n\simeq\delta_N+\omega_N'n, \quad\quad
	\delta_N=\omega_N-\omega_N'N
\end{equation}
where we have used a prime to denote the first derivative, namely $\omega_n'=d\omega_n/dn$. In the next section, we will show that,
despite the nonlinear form of the interaction, a linear behavior of
the eigenfrequencies is indeed possible in some models at least in an energy interval. 

Substituting the  expressions of Eq.~\eqref{eq:taylor} in Eq.~\eqref{eq:Cs} and using the result in \eqref{eq:Psit}
one can approximate the state vector $\ket{\Psi(t)}\simeq\ket{\Psi_{\rm ap}(t)}$ with
following expansion in terms of coherent states of the field
and material Bell states
\begin{align}
\label{eq:PsiFinal}
\ket{\Psi_{\rm ap}(t)}&=
	\left[\ket\zeta
\ket{\alpha}
+\ket{\Upsilon(t)}\right]/\mathcal{N}(t),
\end{align}
where we have identified a time independent contribution, $\ket\zeta\ket\alpha$, with the atomic stationary state
\begin{equation}
	\ket\zeta=c_-\ket{\Psi^-}+d_-\ket{\Phi^-}.
\end{equation}
The time dependence is then present only in the following atoms-oscillator state
\begin{equation}
\ket{\Upsilon(t)}=\textstyle\sum_{\pm}
	b_\pm
	e^{\mp i(\delta_N+S_z\omega_N')t}
	\ket{\phi_\pm}\ket{\alpha e^{\mp i\omega_N' t}},
\end{equation}
that is given in terms of two time dependent coherent states accompanied by the following normalized material states and their initial probability amplitudes
\begin{align}
\ket{\phi_{\pm }}&=
\frac{1}{\sqrt 2}
\left(\ket{\Psi^{+}}\pm\ket{\Phi^{+}}\right),\quad 
b_{\pm }=
\frac{c_{+}\pm d_+}{\sqrt 2}.
\end{align}
Due to the performed approximations, one has to consider the following
normalization 
\begin{align}
	\label{eq:nomalization}
	&
	{\mathcal N}(t)=1+2{\rm Re}[b_{+}^\ast b_{-} e^{i2h(t)}]e^{-2N\sin^2\omega_N' t}\sin^2\omega_N't
	\\
	&-\sqrt2{\rm Im}\left[
	b_{+}e^{-ih(t)}
	+b_{-}e^{ih(t)}
	\right]e^{-2N\sin^2\omega_N' t/2}\sin\omega_N't,
	\nonumber
	\end{align}
with $h(t)=\delta_Nt+N\sin\omega_N't$. This normalization will not play a role in the forthcoming analysis for two reasons. It approaches the unit value
in the limit $N\to\infty$ as can be seen from the behavior $e^{-2N\sin^2 \omega_N't/2}\sin\omega_N't$. The second reason is that it attains unit value whenever
$\omega'_Nt$ is an integer multiple of $2\pi$ and  those will be the interaction times that will draw our attention.  

The approximation in Eq. \eqref{eq:taylor} is valid as long as contributions to the time-evolution corresponding to higher orders in the Taylor expansion of the eigenfrequencies remain negligible. These contributions
have  the form $t\omega_N^{(j)}(n-N)^2/j!$ and can 
be neglected for small values of $t$. 
However, as time elapses, each contribution can be important taking into account that $\omega_n t$ is evaluated inside an exponential
as $\exp(i\omega_n t)$
where its value is taken modulo $2\pi$.
This imposes a restriction on the maximum interaction time
 $t\ll t_{\rm b}$, i.e., when it is considerable less than a breakdown time $t_{\rm b}$, that can be 
 obtained from the condition 
\begin{equation}
	\label{eq:breakdown}
	\frac{|\omega_N^{(j)}(n-N)^jt_{\rm b}|}{j!}
	= 1
	\quad \Rightarrow\quad
	t_{\rm b}=\frac{j!}{(8N)^{j/2}|\omega_N^{(j)}|},
\end{equation}
where $\omega_N^{(j)}$ is the first nonzero derivative of order $j>1$.  In the previous expression we have taken  
into account the standard deviation of the Poissonian distribution given by $\alpha=\sqrt{N}$ and therefore
we have replaced $|n-N|$ with $\sqrt{8N}$. In this way, the sum of $p_n^2$ in the interval
$(N-\sqrt{8N},N+\sqrt{8N})$ is larger than $0.995$.

The result in Eq. \eqref{eq:PsiFinal} is the first important result of this work, as it gives a general expression
of the state vector  for an initial coherent state of the oscillator and arbitrary atomic states. It is to be noted that with the inverse transformation in Eq. \eqref{eq:intpic} one can obtain the state vector in the Schr\"odinger picture in simple way from our final state vector in Eq. \eqref{eq:PsiFinal}. More general states of the field could be eventually considered using our result together with the 
coherent states completeness relation.  
It is also worth mentioning that the present treatment is also feasible in the case of more atoms or multilevel atoms. In these cases, the difficulty lies in the diagonalization of larger blocks of the interaction Hamiltonian that could be achieved in an approximate fashion in order to obtain an expansion in terms of coherent states.
 Similar analytical expressions have been found
for the two-atom Tavis-Cummings model \cite{Jarvis2009a,Torres2014} and first for the Jaynes-Cummings model \cite{Gea-Banacloche1991a}. However, here we have 
presented a more general expression that is valid for any model described by a Hamiltonian of the form of Eq. \eqref{eq:Hamiltonian}. Furthermore,
we will show that with this expression in terms of material Bell sates, it is possible to analyze in a more manageable way the entanglement in the system. 

\subsection{Rabi oscillations and relevant time scales}
\label{sec:Model:Rabi}
Relevant timescales can be unveiled  by evaluating expectation values of the system observables.
It is not hard to realize that these quantities depend on the overlaps between coherent states of the form
\begin{equation}
	\label{eq:overlap}
\braket{\alpha}{\alpha e^{i\omega_N't}}=e^{iN\sin\omega_N' t}e^{-2N\sin^2\omega_N't/2}.
\end{equation}
Let us 
consider, as a figure of merit, the expectation value of $S_z$, Eq. \eqref{eq:atomoperators}, with an initial state  $\ket{\rm ee}\ket\alpha$. Using the overlap between coherent states and
the solution to the time-dependent state vector, Eq. \eqref{eq:PsiFinal}, one can arrive to the approximate  expression
\begin{equation}
	\label{eq:Szt}
	\langle S_z(t)\rangle \simeq 	e^{-2N\sin^2 \omega'_Nt/2}\cos(\delta_N t+N\sin\omega_N't).
\end{equation}
From this expression, one can identify three different time scales. 
The fastest one is given by the Rabi frequency 
$\omega_N$ determining fast oscillatory behavior. The oscillations eventually vanish as they are modulated by a Gaussian envelope; 
a phenomenon known as collapse of Rabi oscillations \cite{Eberly1980,Gea-Banacloche1991a,Torres2014,Jarvis2009a}.
This happens for times with vanishing small values of the exponential in Eq. \eqref{eq:Szt}, when its argument differs from integer (zero included) multiples of $2\pi$.
The oscillations reappear when the argument of the exponential in Eq. \eqref{eq:Szt} vanishes, what is known as revival of Rabi oscillations.
These relevant times can be evaluated from the previous expression and result in 
 the expressions for the Rabi time, collapse time, and revival time that correspondingly are given by
\begin{equation}
	\label{eq:times}
t_{\rm R}=2\pi/\omega_N,\quad
t_{\rm c}=2/\sqrt N|\omega_N'|,\quad t_{\rm r}=2\pi/|\omega_{N}'|.
\end{equation}
One can note that the revival time always scales with the collapse time as $t_{\rm r}=\pi \sqrt N t_{\rm c}$, where $N$ is the mean number of quanta in the oscillator. 
In Sec.~\ref{Dynamical} we will numerically study this behavior for the specific models that will be presented in Sec.~\ref{Specific}.
It is worth commenting that the expression in Eq.~\eqref{eq:Szt} is only valid for times where the linearization in Eq.~\eqref{eq:taylor} represents a faithful approximation
of the eigenfrequencies $\omega_n$. 

\subsection{State vector at fractional revival times}
The revival time $t_{\rm r}$, as previously introduced in Eq.~\eqref{eq:times}, corresponds to the moment at which all components of the oscillator state  in Eq.~\eqref{eq:PsiFinal}
return to the initial the condition $\ket\alpha$.  At fractional multiples of this revival time, the complete system attains interesting and relevant states
\cite{Jarvis2009a,Torres2014,Torres2016b}. For instance, 
the state vector at each odd integer multiple of a quarter of the revival time, $t_{\rm r}/4$, is  given as the completely separable state 
\begin{align}
	\label{eq:Psipi4}
	\ket{\Upsilon\left(\tfrac{kt_{\rm r}}{4}\right)}&=
\ket{\zeta_{1,k}}\,
	{\textstyle\sum_{\pm}}
r^{-1}b_\pm e^{\mp ik\delta_N t_{\rm r}/4}
	\ket{\mp i \alpha},\\	
	\ket{\zeta_{1,k}}&=r\frac{\ket{\Psi^+}+i^{k}\ket{\Phi^-}}{\sqrt 2},\quad r=\sqrt{|c_+|^2+|d_+|^2},
	\nonumber
\end{align}
with an odd integer $k$. We have arrived to this state using Eq. \eqref{eq:PsiFinal} and the relation
\begin{equation}
	\label{eq:EZPi2}
	e^{\mp i S_z\pi/2}\ket{\Phi^+}=\pm i\ket{\Phi^-}.
\end{equation}
It can be noted that in the state of Eq. \eqref{eq:Psipi4}, matter and oscillator separate and that the atomic state is independent from the initial condition. Perhaps not so evident is the fact
that the atomic state is a separable state for any value of $k$, a property that can be simply tested with any entanglement measure, such as the concurrence
that will later be used in this work. This means that even if the atoms where  initially entangled, no entanglement remains at this time in any partition of the systems such as:
atom-atom or (any atom or both atoms)-field.  This phenomenon, with no entanglement in the system even if it was initially entangled, 
has been refereed to as ``basin of attraction'' in the Tavis-Cummings model \cite{Jarvis2009a}. It is important to note that this only happens for the 
time-dependent part of the state, $\ket{\Upsilon(t)}$, and therefore, this feature applies only when the stationary part vanishes, i.e., whenever $c_-=d_-=0$.

At odd multiples of one half of the revival time, the time-dependent part is given by
\begin{equation}
\label{eq:Psipi2}
\ket{\Upsilon(\tfrac{k}{2}t_{\rm r})}=
\ket{\zeta_{2,k}}\ket{-\alpha},\quad \ket{\zeta_{2,k}}=c_{k}\ket{\Psi^+}+d_{k}\ket{\Phi^+}
\end{equation}
with an odd integer $k$ and the coefficients given by
\begin{align}
	\label{eq:ckdk}
	c_k&=c_+\cos \delta_N \frac{k}{2}t_{\rm r}  -i d_+\sin \delta_N \frac{k}{2} t_{\rm r}\\
	d_k&=-i^{2k+1}c_+\sin \delta_N \frac{k}{2}t_{\rm r} + i^{2k}d_+\cos \delta_N \frac{k}{2} t_{\rm r}.
\end{align}
In this case, one has again a product state of atoms and oscillator. However, in this case,
the atomic part might be entangled. It is not hard to realize, as we will later show, that $\ket{\zeta_{2,k}}$ has the same degree of entanglement as the initial component
$c_{+}\ket{\Psi^+}+d_{+}\ket{\Phi^+}$. For this reason, Eq. \eqref{eq:Psipi2} will play an important role in identifying the entanglement properties in the system
and in order to design entangling operations that  will be shown in Sec.~\ref{sec:EntanglingOperations}.

\section{Specific models}
\label{Specific}
\label{sec:models}
In this section we present three examples of models that can be described by the interaction Hamiltonian in Eq.~\eqref{eq:V}. We start with the Tavis-Cummings model in order
to compare our results with the most studied example \cite{Jarvis2009a,Tessier2003,Torres2014}.  
The Buck-Sukumar model \cite{Buck1981} is considered as it presents
a particular nonlinear interaction that induces an
almost exact linear behavior of the eigenfrequencies as required in Eq.~\eqref{eq:taylor}. 
An ion-trap nonlinear
model \cite{Vogel1995} will be considered, as it represents a viable experimental setting to this problem. We will demonstrate that, despite the intrinsic nonlinear behavior, a linearization of the eigenfrequencies is possible in a restricted interval of the oscillator occupation number. 
\subsection{Two-atom Tavis-Cummings model}
The Tavis-Cummings model describes the interaction of an arbitrary number of two-level atoms interacting with a single-mode of the quantized electromagnetic field  \cite{Tavis1968b}. 
It can be viewed as an extension of the Jaynes-Cummings model \cite{Jaynes1963} for many atoms and it has therefore become
a paradigm in cavity QED. The original model was introduced in the same form as in Eq. \eqref{eq:Hamiltonian} with $f(a^\dagger a)=1$ and with pseudo momentum
operators $S_\pm$ and $S_z$ for arbitrary number of two level particles. Here, however, we only consider the two-atom case that corresponds to the atomic operators
in Eq. \eqref{eq:atomoperators} and whose interaction Hamiltonian is diagonalizable in the block form of Eq. \eqref{eq:BlockV}.

As in this case $f(a^\dagger a)=1$ in Eq. \eqref{eq:V}, the matrix elements in the 
blocks of the interaction potential, Eq. \eqref{eq:BlockV}, can be obtained from  $\Omega_n=\Omega \sqrt{2n+2}$. The eigenfrequencies or Rabi frequencies 
are obtained from Eq. \eqref{eq:omegaeig} and are	$\omega_n=\Omega\sqrt{4n+2}$. 
The relevant frequencies determining the total state in Eq. \eqref{eq:PsiFinal} can be found using Eq. \eqref{eq:taylor} as
\begin{equation}
\label{omegasTC}
	\omega_N'=\frac{2\Omega}{\sqrt{4N+2}},\quad \delta_{N}=\frac{2N+2}{\sqrt{4N+2}}.
\end{equation}
Therefore, in this model one can find that the relevant time scales are given by
\begin{equation}
	t_{\rm R}\approx \frac{2\pi}{\Omega\sqrt N} ,\quad  t_{\rm b} \approx \frac{\sqrt N}{\Omega},\quad  t_{\rm r}=\frac{2\pi\sqrt N}{\Omega}.
\end{equation}
The shortest time scale corresponds to the Rabi oscillations period $t_{\rm R}$, followed by 
the time $t_{\rm b}$ when the coherent state approximation breaks down, see Eq. \eqref{eq:breakdown}. Finally one has the reappearance of Rabi oscillations
at the revival time $t_{\rm r}$.  As $t_{\rm r}>t_{\rm b}$, the revival of Rabi oscillations is not perfect in the Tavis-Cummings model and for this reason the field components will
deform leading to the well known broadening of the revivals \cite{Eberly1980}.

\subsection{Two-atom Buck-Sukumar model}
In 1980 Buck and Sukumar presented a simple theoretical model 
for the interaction of a two-level atom with a single-mode electromagnetic field \cite{Buck1981}. 
In this model the atom-field coupling is assumed 
to be nonlinear in the field variables and 
can be interpreted as an intensity-dependent interaction. As the Buck-Sukumar model (BS) 
is integrable and allows perfect revivals 
of Rabi oscillations in the case of initial coherent fields, it has drawn considerable theoretical attention in the past \cite{Kochetov1987,Bonatsos1993,Rybin1999}.
A drawback of this model, however, is that there is no obvious physical implementation. 

Here we consider the Buck-Sukumar interaction for the two-atom case, where  $f(a^\dagger a)=\sqrt{a^\dagger a}$ in Eq. \eqref{eq:V}. This implies a linear
dependence on $n$ in the matrix elements of the blocks of $V$ and its eigenfrequencies, namely $\Omega_n=\Omega\sqrt2 (n+1)$ and $\omega_n=(2n+1)\Omega$. The relevant frequencies in the time-dependent state vector in \eqref{eq:PsiFinal} are simply given by
\begin{align}
\label{omegasBS}
\omega_N'=2\Omega,\quad \delta_N=\Omega.
\end{align}
The timescales are dictated in this case by the following parameters
\begin{equation}
	t_{\rm R}\approx \frac{\pi}{\Omega N},\quad  t_{\rm r}=\frac{\pi}{\Omega},\quad t_{\rm b} =\frac{N^2}{\sqrt2\Omega}.
\end{equation}
In contrast to the Tavis-Cummings model, here the breakdown time  of the coherent state approximation $t_{\rm b}$ scales as $N^2$. In this case,
the approximate value of the eigenfrequencies are linear with $n$ and therefore predict an infinite value of $t_{\rm b}$. Therefore, 
we have used the exact dependence on $n$ of the eigenvalues, which is $\Omega\sqrt{4n^2+4n+2}\simeq(2n+1)\Omega$. 
 Another important difference is that here the revival time is independent of  the mean value of the oscillator $N$.

\subsection{Ion-trap nonlinear model}
\begin{figure}
	\includegraphics[width=0.23\textwidth]{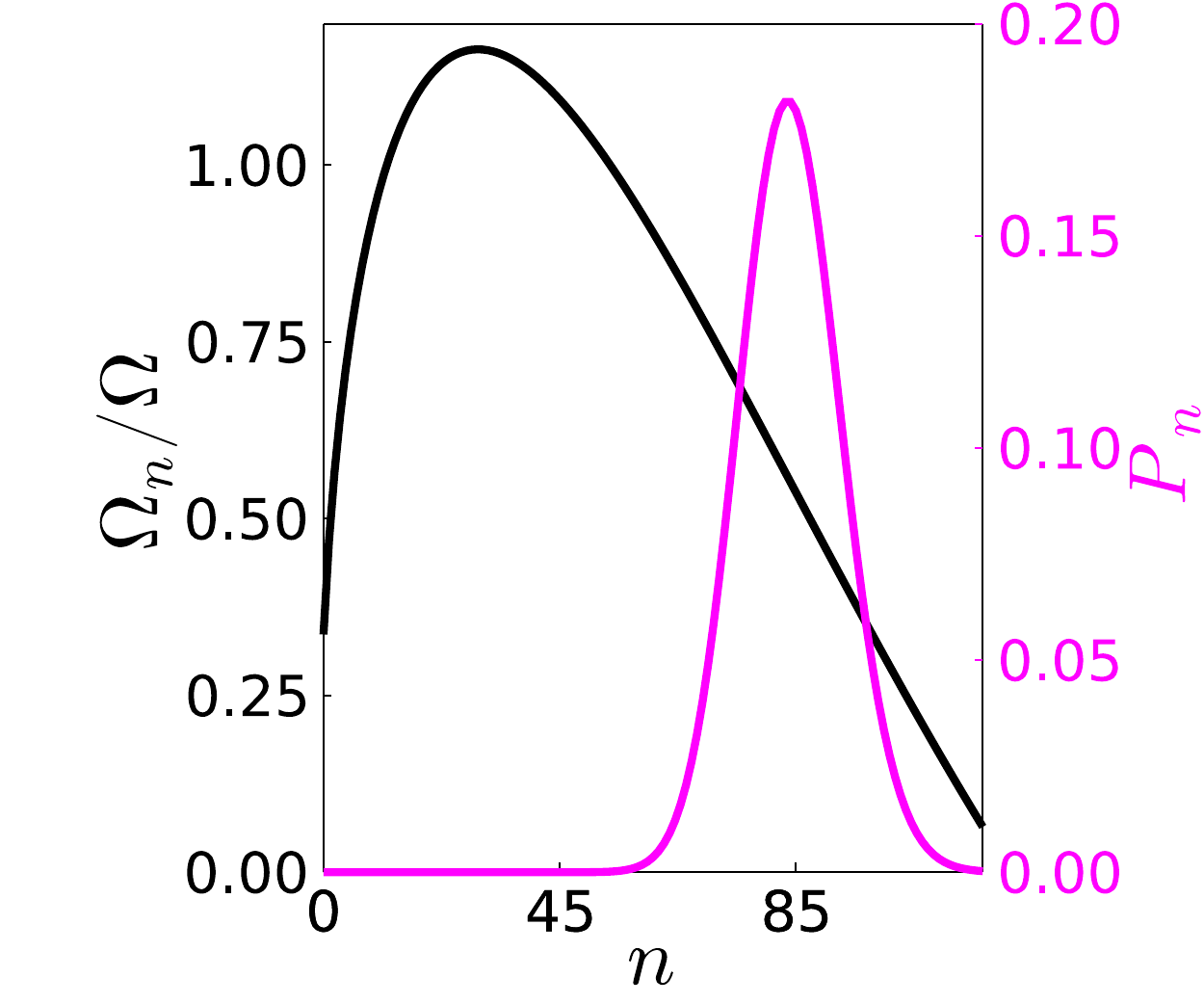}
		\includegraphics[width=0.23\textwidth]{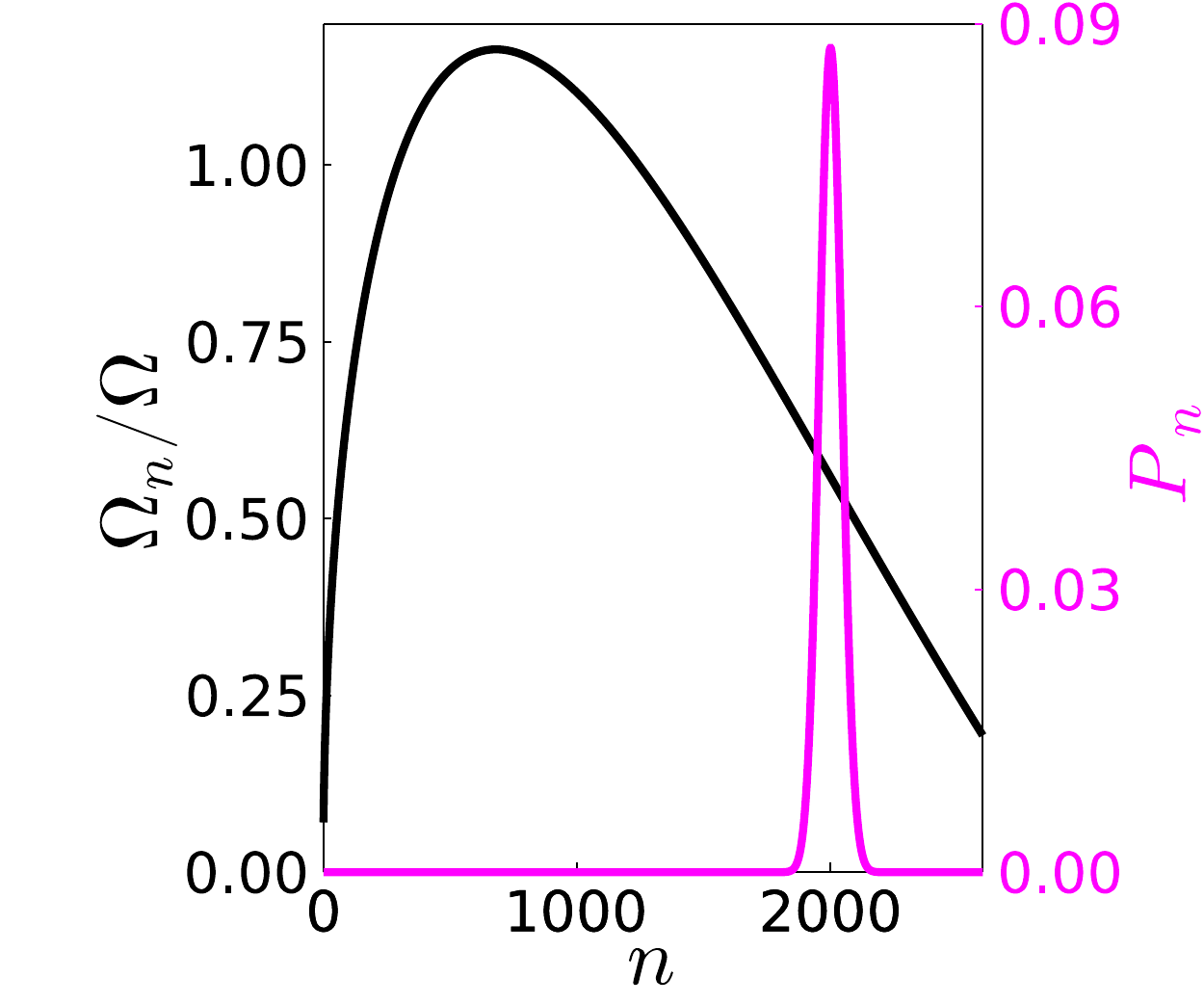}
	\caption{\label{fig:WnPn}Eigenfrequency $\Omega_n$ as a function 
	of the oscillator quantum number $n$ for two different values of the mean number of quanta and the Lamb-Dicke parameter: $N=85$,  
	$\eta=0.170582$ (left), and $N=2000$, $\eta=0.0352653$ (right). Approximately linear
	behavior can be appreciated around $N$. In magenta (light gray line), we present the probability amplitude $p_n$ of a number state in a coherent state $\ket\alpha$, with $N=\alpha^2$.
	In both cases, vanishing small contributions of $p_n$ lie outside of the apparent linear interval of $\Omega_n$.
}
\end{figure}

The last and most important model that will be considered consists on two ions trapped in a linear harmonic potential driven by a 
classical monochromatic radiation field. In this case $a$ and $a^\dagger$ 
represent the annihilation and creation operators of the ions center of mass motion \cite{Vogel1995,Roos2008}. 
 The free Hamiltonian is given by $H_0=\hbar \omega S_z+\hbar\nu a^\dagger a$, i.e., 
the frequency $\nu$ of the mechanical oscillator differs from the transition frequency of the atoms. 
The coupling with the electronic levels is mediated by the external monochromatic field
whose frequency is tuned to the first vibrational sideband and is given by $\omega_{\rm L}=\omega-\nu$. 
With these conditions, the interaction Hamiltonian is time independent in the interaction picture and  is also well described by Eq.~\eqref{eq:V}
with the following intensity-dependent function
\begin{equation}
\label{nTC:Hamiltonian}
f(a^{\dagger}a)=\eta e^{-\eta^2/2}\sum_{m=0}^{\infty}\frac{(-\eta^2)^m}{m!(m+1)!}a^{\dagger m}a^{m}.
\end{equation}
Details of the derivation are given in Appendix \ref{iontrapV}.
In this case, the nonzero matrix elements of the interaction potential can be expressed in terms of a Laguerre polynomial, namely
\begin{equation}
\label{eq:OmegaLagguerre}
	\Omega_n=\Omega\eta\sqrt{\frac{2}{n+1}} e^{-\eta^2/2}L_n^{(1)}(\eta^2).
\end{equation}
This polynomial will clearly display nonlinear behavior
that will be inherited by the eigenfrequencies  $\omega_n$. However, for a given value
of the Lamb-Dicke paramenter $\eta$, it is possible to find an interval
around a certain value of $N$ displaying approximately linear behavior with $n$. In principle, it is possible to find the most suitable value of the mean phonon number 
for a given value of the Lamb-Dicke paramenter $\eta$ by analyzing the form
of the Laguerre polynomial as a function of $n$. However, the task greatly simplifies 
by expressing the Laguerre polynomials in terms of Bessel functions 
\cite{Szego1975,Muckenhoupt1970,Alonso2016} which, in our case, is a good approximation
whenever $\eta^2\ll 4n+4$.
Doing so one can find the following approximate expression $\Omega_n\simeq \sqrt{2}\Omega J_1(2\eta\sqrt{n+1})$ and therefore the eigenfrequencies become
\begin{equation}
	\label{eq:omegaIT}
\omega_n\simeq 2\Omega\left|J_1\left(2\eta\sqrt{n+\tfrac{1}{2}}\right)\right|,
\end{equation}
where $J_1(\sqrt{x})$ is the Bessel function of first kind and order one. 
The eigenfrequency $\omega_n$  is plotted 
in Fig.~\ref{fig:WnPn} for two different values of the Lamb-Dicke paramenter. Relating the argument of the Bessel function as $ 2\eta\sqrt{n+1/2}=\sqrt x$,
it is possible to analyze the function for arbitrary values of $\eta$. One can then
note that there is an approximate linear behavior in
the interval $x\in(7.25,12.65)$. Indeed, one can realize that a linear approximation in this
interval differs on average from the original function in less than  $1\%$. For this estimation, we have performed a Taylor expansion around $x_0$, the zero  of the function $d^2J_1(\sqrt x)/dx^2$, which is the point where the slope of $J_1(\sqrt x)$ changes behavior.
In this way, one is able to
find a relation between the mean number of quanta $N$ and the Lamb-Dicke paramenter $\eta$ as
\begin{align}
	\label{eq:eta}
    N=&\frac{x_0}{4\eta^2}-\frac{1}{2},\quad x_0=9.95161.
\end{align}
The value of $x_0$ is written to six digits precision and it
was obtained using the Newton-Raphson method.
The value of  $N$ decreases as $\eta$ 
increases. Therefore, 
in order to fit a Poissonian distribution
with standard deviation $\sqrt N$
 in the linear interval,
one  has to fulfill the condition $\eta\le 2.7/\sqrt{32x_0}\approx 0.156905$. For this reason, large values of the Lamb-Dicke parameter cannot be used in this scheme. 
In Fig.~\ref{fig:WnPn} we have also plotted the probability amplitude of each number state in the coherent state of Eq. \eqref{eq:cohstate} for two different values of 
the mean number of quanta $N$. It is to be noted that for a smaller value of the Lamb-Dicke paramenter, the mean number $N$ increases and also the number
of states lying in the linear part of the function. For this reason, in the limit of large $N$, one does not require to perfectly fit the optimal value of $N$ in
Eq.~\eqref{eq:eta}. The generation of large motional coherent states in trapped ions is nowadays possible \cite{Alonso2016,Johnson2017} offering interesting perspective to implement
this model.

Using the results in Eqs. \eqref{eq:omegaIT} and \eqref{eq:eta} one can obtain the relevant frequencies for the state vector \eqref{eq:PsiFinal} as 
\begin{align}
	\label{eq:ITvalues}
	\omega_N&=2\Omega J_1(\sqrt{x_0})\approx0.558924\Omega\nonumber\\
    \omega_N'&=\frac{\Omega\sqrt{x_0}}{2N+1}(J_0(\sqrt{ x_0})-J_2(\sqrt{ x_0}))\approx
    -\frac{2.50163\Omega}{2N+1}\\
    \delta_N&=\omega_N-N\omega_N'\approx 
    \left(
     0.558924+\frac{2.50163N}{2N+1}.
     \right)\Omega
     \nonumber
\end{align}
An important feature to note here is that this quantities are given only in terms of the optimal  value of $N$, therefore, indirectly depending on $\eta$. In this form,
a similar analysis as for the previous two models is also possible in this case. As for the time scales, it is no difficult to find that the relevant values are given by
\begin{equation}
	t_{\rm R}\approx \frac{2\pi}{0.56\Omega} ,\quad 
  t_{\rm r}=\pi\frac{4 N+2}{2.5\Omega},\quad t_{\rm b} \approx \frac{N^{3/2}}{10\Omega}.
\end{equation}
In this model, the period of the Rabi oscillations is independent of the mean phonon number $N$ and the revival time $t_{\rm r}$ scales linearly with $N$.
The breakdown time of the coherent state approximation roughly relates to the revival time as $t_{\rm b}\approx \sqrt N/50$. Therefore, in order to have a faithful
description, one has, in principle, to achieve large mean phonon numbers. For instance, for an accurate description up to an interaction time $t_{\rm r}/2$ one requires 
values of $N>625$. In the next section, however, we will show that even with moderate values of $N$, the model offers a reasonable  description. 

In order to present a clear comparison between the models, in Table \ref{tab:1} we present a summary of the dependence on $N$ of the different times for the three cases presented.
\begin{table}[ht]
 	\centering
 		\begin{tabular}{c c c c c}
 			\multicolumn{4}{c}{} \\ \hline
 			\textbf{Model}& $\Omega t_{\rm R}$&$\Omega t_{\rm c}$&$\Omega t_{\rm r}$ & $\Omega t_{\rm b}$\\ \hline
 Tavis-Cummings & $2\pi/\sqrt N$ & 2 & $2\pi\sqrt{N}$ & $\sqrt N$ \\ \hline	
 Buck-Sukumar & $\pi/N$ & $1/\sqrt{N}$ & $\pi$ & $N^2/\sqrt2$ \\ \hline	
 ion-trap & $11.2$ & $1.6\sqrt{N}$ & $5 N$ & $0.1 N^{3/2}$ \\ \hline	
 		\end{tabular}
 		 \caption{\label{tab:1} Relevant time scales for three different models in terms of the mean number of photons $N$: 
 		Rabi oscillations period, collapse time, revival time, and breakdown time of the coherent state approximation. }
 		
 \end{table}

\section{Dynamical features}
\label{Dynamical}
In this section we present the results and comparison of numerical calculations of
 dynamical features of the three specific models introduced in Sec. \ref{sec:models}. We focus on the collapse and revival of Rabi oscillations and we test our analytical result 
 with numerically exact calculations that are evaluated
 using the exact form of the state vector in Eq. \eqref{eq:Psit}
 using the expressions in Appendix \ref{Exact}.

\subsection{Rabi oscillations and phase space representation}
\begin{figure}[t]
	\includegraphics[width=0.45\textwidth]{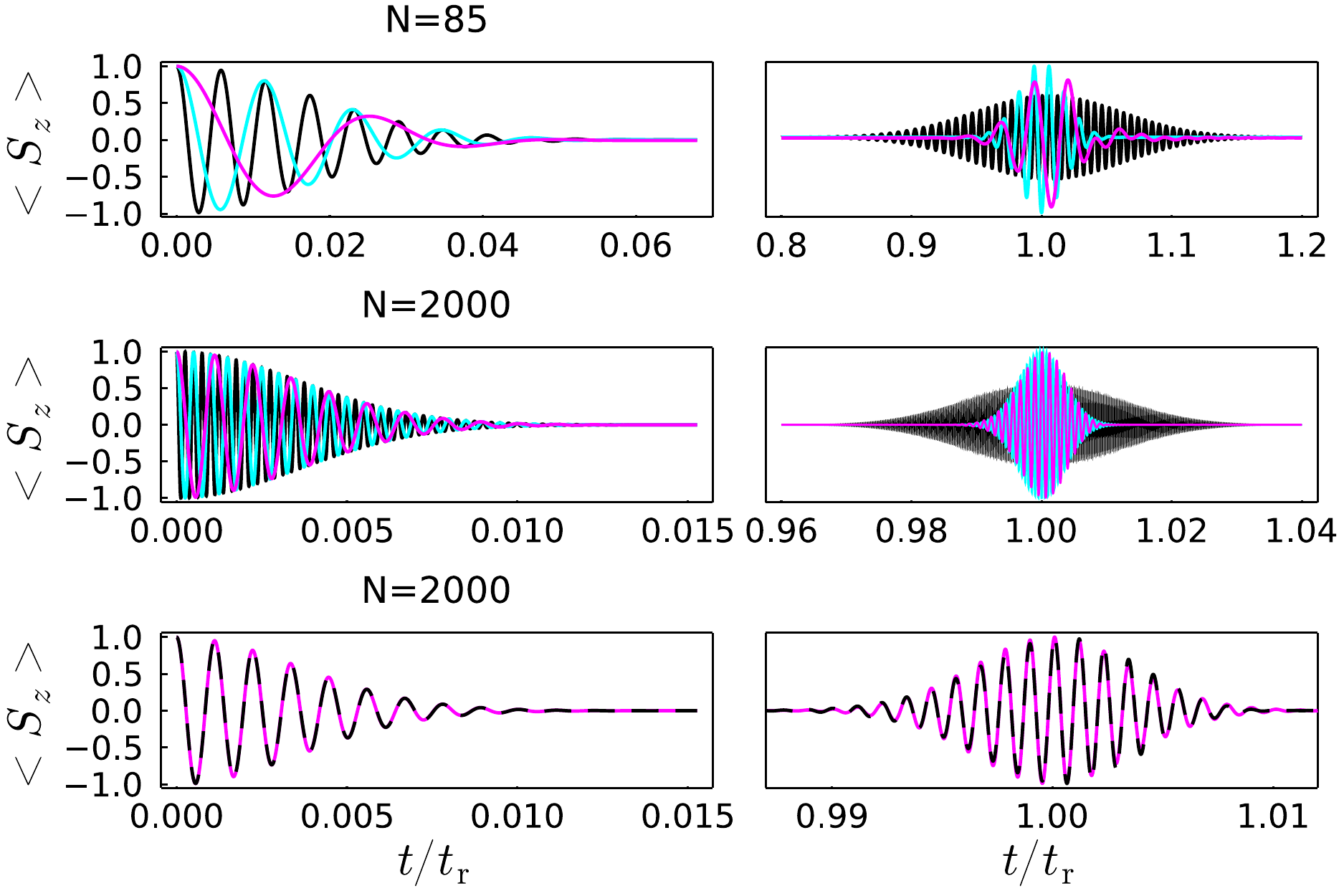}
	\caption{\label{fig:RabiOsc}Expectation value of the operator $S_z$ in Eq. \eqref{eq:atomoperators} for two different values of the mean number of quanta: $N=85$ (top plots), $N=2000$ (middle plots).
		In the left column, the initial Rabi oscillations and its collapse is presented. In the right column the first revival of Rabi oscillations is displayed around
		a time $t_{\rm r}$. Black, cyan (light gray), and magenta (gray) curves respectively correspond to the Tavis-Cummings model, the Buck-Sukumar model, and the ion-trap model. In the bottom plots we present a comparison between the analytical prediction in Eq. \eqref{eq:Szt}  (dashed curve) and the numerically exact calculation (solid line) using the ion-trap model conditions with $N=2000$.}
\end{figure}

As mentioned in Sec. \ref{sec:Model:Rabi}, the relevant time scales of the system can be obtained from evaluating the expectation value of observables in the system. 
As figure of merit, in this work we have chosen to evaluate the mean value of $S_z$, which can be analytically evaluated from our approximate expression 
in \eqref{eq:PsiFinal} with the result given in \eqref{eq:Szt}.	In Fig.~\ref{fig:RabiOsc} we have plotted the numerically exact result of $\langle S_z(t)\rangle$ 
for the different models with
an initial atomic state $\ket{\rm ee}$ and mean number of photons $N=85$ and $N=2000$. 
The black, cyan (light gray), and magenta (gray) curves correspond, respectively, to the Tavis-Cummings,
Buck-Sukumar and ion-trap models.
 We present the Rabi oscillations close to $t=0$ (left column) and around $t=t_{\rm r}$
(right column).
The first evident feature is that for all three models the collapse of the Rabi oscillations occurs at the same fraction of the revival time $t_{\rm r}$, i.e., the Gaussian
envelope is the same in terms of the adimensional time $t/t_{\rm r}$.  This is in complete agreement with the analytical approximation given in Eq.~\eqref{eq:Szt}. For the
reappearance of the Rabi oscillations around $t_{\rm r}$, only the Tavis-Cummings model presents a broadening of the oscillatory region. The Buck-Sukumar model presents perfect 
revivals for the two values of $N$. The ion-trap model presents no apparent enhancement, however, for $N=85$ the oscillations display asymmetries. The revival seems to be perfect 
in this model for $N=2000$.

The collapse and revival of Rabi oscillations can be elucidated by visualizing the state of the oscillator in phase space
 with the aid of some quasiprobability distribution. In this work we rely on the Husimi function 
 that can be regarded as the expectation value of the oscillator reduced density matrix $\rho_{\rm os}$ with respect to a coherent state $\ket\beta$, namely
 \begin{align}
 	Q(\beta)&=\bra\beta \rho_{\rm os}(t)\ket\beta/\pi, 
 	\\ \rho_{\rm os}(t)&={\rm Tr}_{\rm at}\left\{\ketbra{\Psi(t)}{\Psi(t)}\right\}.
 	\nonumber
 \end{align}
We have used the notation ${\rm Tr_{at}}$ for the partial trace with respect to the atomic electronic degrees of freedom and we have considered $\beta$ as a complex parameter.
Reconstruction of a Husimi $Q$-function has been experimentally achieved on single ${}^{171}\rm{Yb}^{+}$ ions in a harmonic potential by using Raman laser beams \cite{Dingshun2017}.

\begin{figure}[t]
	\centering
	\includegraphics[width=0.45\textwidth]{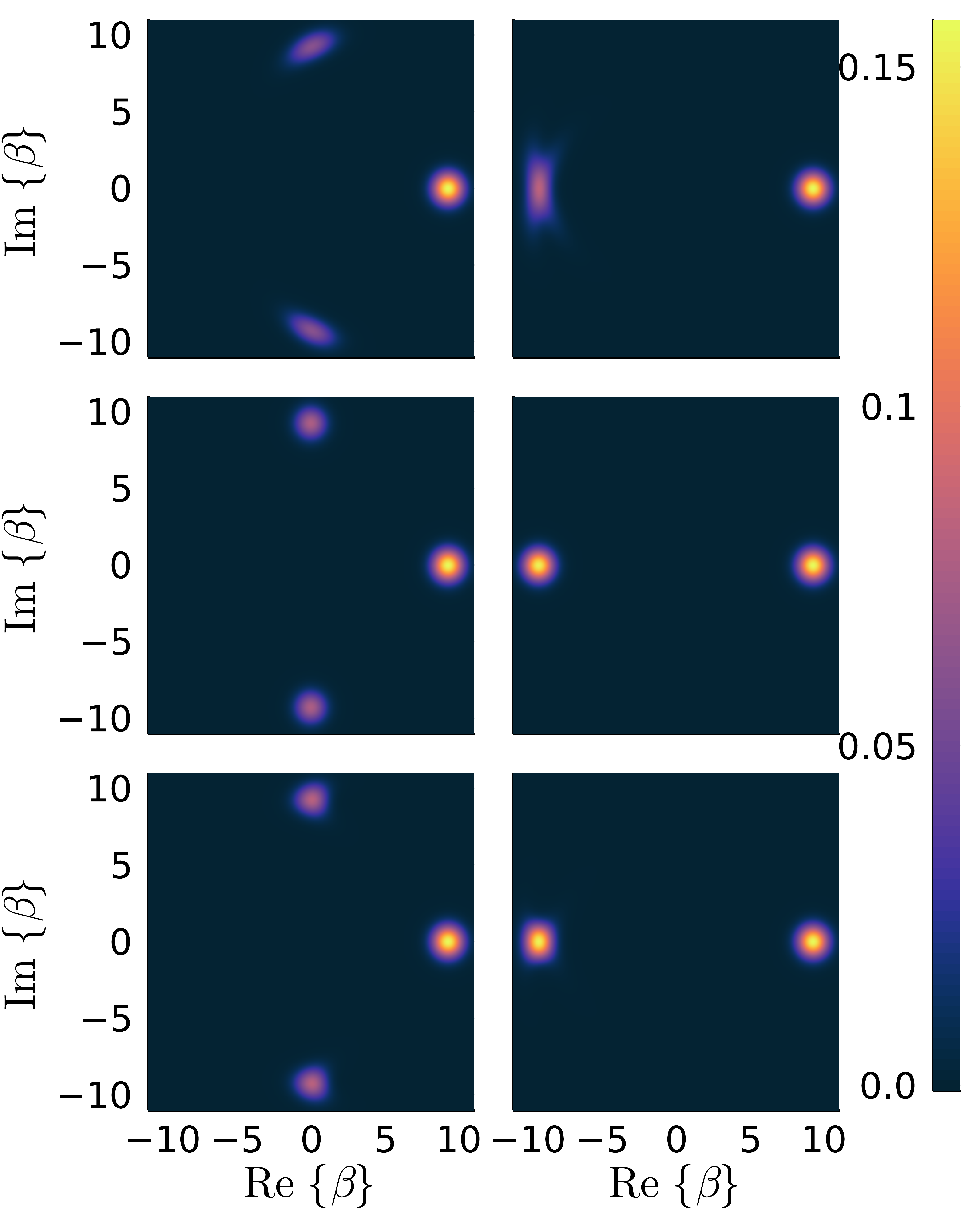}
	\caption{
		\label{fig:husimi}
		Husimi function of the reduced density matrix for the oscillator in an initial coherent state $\ket\alpha$ and for interaction times $t=r_{\rm r}/4$ (left column) and
		$t=t_{\rm r}/2$ (right column). The results corresponds to the Tavis-Cumming model, the Buck-Sukumar model, and the ion-trap model for the first, second, and third row respectively. 
	}
\end{figure}

In Fig.~\ref{fig:husimi} we have plotted the Husimi function $Q(\beta)$ for the three models described in Sec. \ref{Specific} and for two different interaction times:
$t_{\rm r}/4$ and $t_{\rm r}/2$. We have used two excited atoms as initial state and a coherent state for the oscillator with $\alpha=\sqrt{85}$. The initial state $\ket\alpha$ remains as a stationary component of the mode for all time as evidenced in the plots. 
It can be noted that the time-evolving field components of the Tavis-Cummings model (top plots) suffer from a distortion already for a time  $t_{\rm r}/4$ and this feature
is more notorious at $t_{\rm r}/2$. In contrast, all mode components in the Buck-Sukumar model (middle plots) retain their shape. This 
an evidence of their evolution as coherent states. In the case of the ion-trap model, the field components follow the same trajectory slightly distorting their shape. This behavior corroborates a good agreement with the coherent state approximation, even with the moderate value $N=85$. 

\subsection{Fidelity of the approximate state vector}
    \begin{figure}[t]
	\centering
	\includegraphics[width=0.42\textwidth]{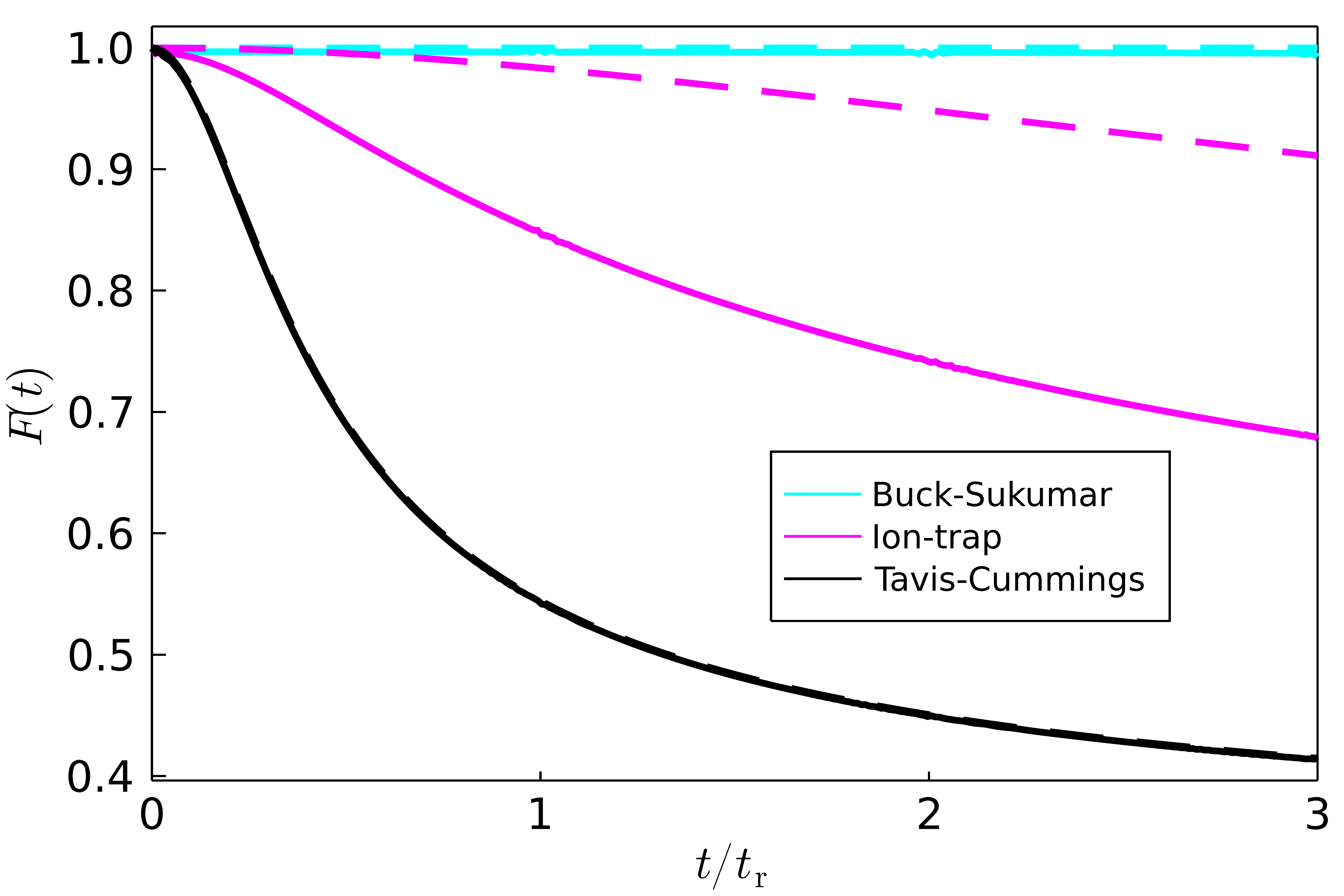}
	\caption{
		\label{fig:fidelity}
		Average fidelity as a function of time of the approximated state vector in \eqref{eq:PsiFinal} with respect the numerically exact state vector
		for two different values of the mean number of quanta: the solid (dashed) line correspond to $N=85$ ($N=2000$). 
		The average has been preformed over $1000$ random initial conditions. 
	}
\end{figure}
In the previous subsection we have briefly analyzed the collapse and revival phenomenon. We have observed that the approximations given in Sec. \ref{Model}
seem plausible given the fact that the revival of the oscillations and the mode components in phase space do not broaden for the Buck-Sukumar and the ion-trap models. 
Let us now turn our attention to the numerical analysis of the validity of our analytical calculation.
In order to test the approximation in Eq. \eqref{eq:PsiFinal}, we consider the fidelity between the exact state vector $\ket{\Psi(t)}$ and its
approximation $\ket{\Psi_{\rm ap}(t)}$ as a function of time that is given by
\begin{equation}
F(t)=|\braket{\Psi_{\rm ap}(t)}{\Psi(t)}|^2/{\mathcal{N}}.
\end{equation}
The normalization $\mathcal{N}$ of $\ket{\Psi_{\rm ap}(t)}$ is given in \eqref{eq:nomalization} and, as mentioned before, it gives only as small contribution close the revivals of
oscillations.
In Fig. \ref{fig:fidelity} we have plotted the fidelity $F(t)$  averaged over $1000$ random initial conditions
 uniformly distributed according to the corresponding Haar measure.
  For the three cases 
we have chosen two different values of the mean number of quanta: $N=85$ presented in full line and $N=2000$ in dashed line. 
For the  Tavis-Cummings model (black curves) the fidelity drops well before the first revival. The Buck-Sukumar model 
(cyan or light gray curves) displays very good fidelity 
for the complete time interval. This is expected as the coherent state approximation is predicted to hold for longer  time, as in this case $t_{\rm b}/t_{\rm r}\propto N^2$. 
For the  ion-trap model (red curves),  the fidelity is maintained above $0.9$ for $N=85$ and greatly improves for $N=2000$, corroborating the expected agreement
given by the time scales in Table \ref{tab:1}.

\begin{figure}[t]
    \centering
    \includegraphics[width=0.47\textwidth]{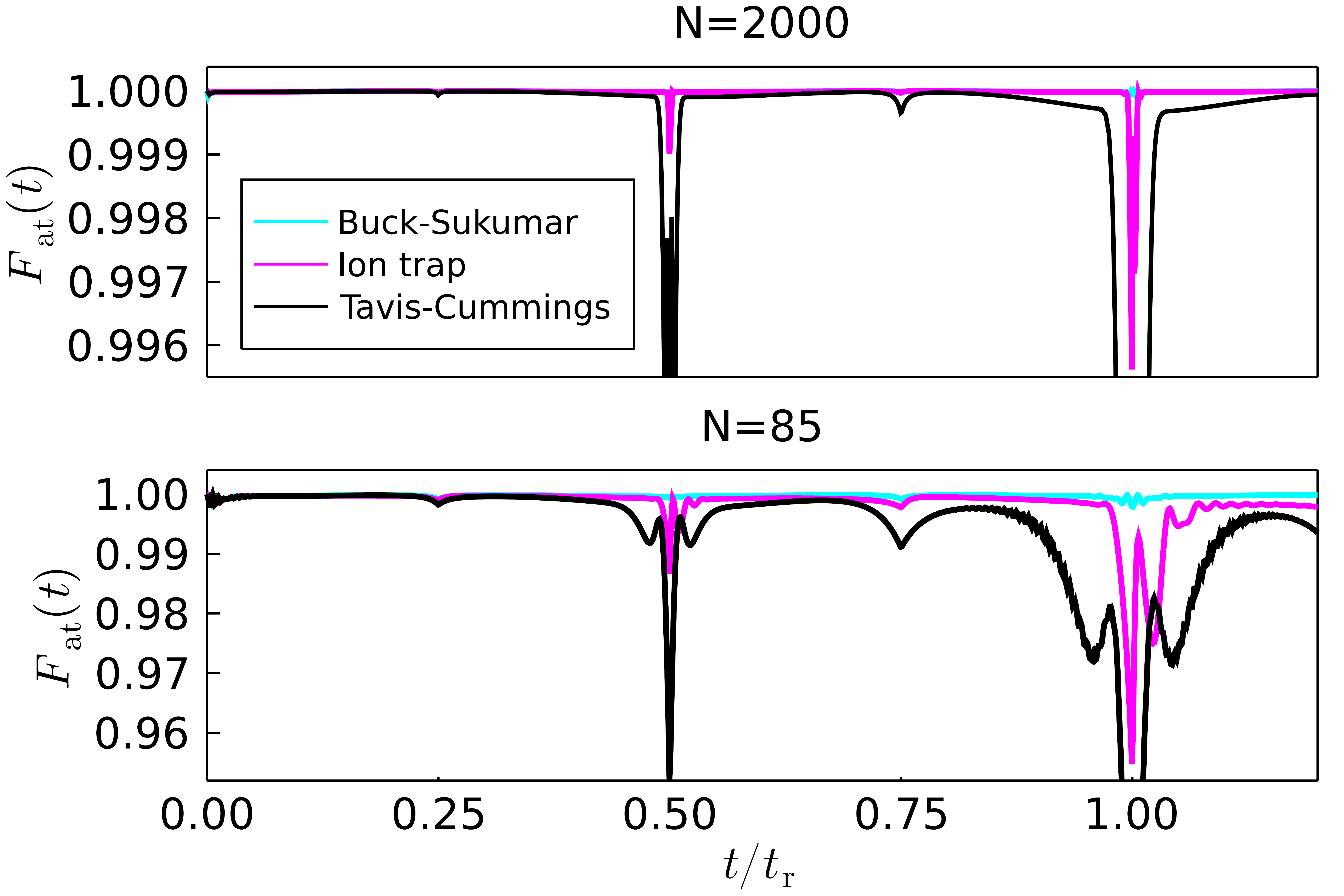}
    \caption{
    	Fidelity of the approximate atomic reduced density matrix with respect to its numerically exact counterpart. 
     averaged over $1000$ initial conditions. Two values of $N$ are considered and indicated in the plot for the three different models as indicated in the legend.
 }
    \label{fig:fidelityAtomic}
    \end{figure}

In the context of quantum computation and quantum information tasks with atomic qubits, the oscillator might be considered as an auxiliary degree of freedom. In this
situation, the state of the mode does not play an important role, and one is mainly concerned with the atomic state. Therefore, the most important state to test
is the reduced density matrix of the atoms whose fidelity with respect to the exact reduced state can be evaluated as
\begin{equation}
	F_{\rm at}(t)=\left({\rm Tr}\sqrt{ \sqrt{\rho_{\rm at}(t)}\rho_{\rm at}^{\rm ap}(t)\sqrt{\rho_{\rm at}(t)}}\right)^2.
\end{equation}
The reduced atomic  density matrices are taken from the exact and approximated total state vector as 
$\rho_{\rm at}^{\rm ap}(t)={\rm Tr}_{\rm osc}\ketbra{\Psi_{\rm ap}(t)}{\Psi_{\rm ap}(t)}$ and similarly
for  $\rho_{\rm at}(t)={\rm Tr}_{\rm osc}\ketbra{\Psi(t)}{\Psi(t)}$.
In  Fig. \ref{fig:fidelityAtomic} we plot the atomic fidelity $F_{\rm at}(t)$ as function of time and  averaged over $1000$ random initial conditions. 
Remarkably, the fidelity is extremely good for all the models, including the Tavis-Cummings model, for times where the field components separate, i.e., for
times different to $t_{\rm r}$ and $t_{\rm r}/2$. Around these times, the Tavis-Cummings model fails to achieve a good fidelity, however, the Buck-Sukumar model displays
good fidelity for any value of $N$. In the case of the ion-trap model, the fidelity increases with $N$. This result also corroborates that the coherent state approximation accurately 
describes the Buck-Sukumar model and the ion-trap model for large values of $N$. In the case of the Tavis-Cummings model, although the coherent state approximation fails
to describe the complete state, the atomic state is well described for times that are not close to the revival time and half the revival time. This happens because the field components follow the trajectory of the coherent states in the approximation. 

\section{Entanglement analysis}
\label{Entanglement}
Entanglement is an important feature of the system in the context of quantum information and quantum computation, especially the atomic entanglement
when the atoms are regarded as qubits. This quantity has been previously studied for the Tavis-Cummings model \cite{Tessier2003,Gonzalez2016} and some interesting properties have been introduced Refs. \cite{Jarvis2009a,Torres2014}. However, 
the quantitative study has been limited to specific initial conditions and numerical calculations.  
As the system is exactly solvable, one could, in principle, calculate in closed form certain entanglement measures 
for any bipartition of the system. However, the resulting expressions will surely be complicated and difficult to analyze. 
Here, we take advantage of our approximation in order to evaluate remarkable simple 
analytical expressions for any initial condition at specific times. In order to carry out this study, we 
evaluate the reduced density matrix for the two-qubit
system given by
$
	\rho_{\rm at}(t)={\rm Tr_{osc}} \ketbra{\Psi(t)}{\Psi(t)}
$.
Furthermore, we concentrate our attention to specific interations times given by $jkt_{\rm r}/4$, namely
at odd multiples $k$ of a quarter (half) of the revival time for $j=1$ ($j=2$). At these times the density matrix assumes the following simple form
\begin{equation}
	\label{eq:rhoat}
	\rho_{\rm at}\left(\tfrac{jk}{4}t_{\rm r}\right)=
	\ketbra{\zeta}{\zeta}+\ketbra{\zeta_{j,k}}{\zeta_{j,k}}
\end{equation}
with $k$ and odd positive integer. For $j=1$ one has to use the state in Eq. \eqref{eq:Psipi4}, and the state in Eq. \eqref{eq:Psipi2} for $j=2$. 
The ket $\ket\zeta$ is the stationary atomic state given in Eq. \eqref{eq:PsiFinal}.
Using $\rho_{\rm at}$ it is possible to
evaluate the entanglement between the atoms and the entanglement between atoms and the oscillator. 

\subsection{Two-atom entanglement}
We rely on the concurrence \cite{Wootters1998a} as a measure of the entanglement between the atoms. For a general two-qubit sate $\rho$, it is defined as
\begin{equation}
	\label{eq:concurrence}
	C(\rho)=\max(0,\lambda_1-\lambda_2-\lambda_3-\lambda_4),
\end{equation} 
where
the four $\lambda_i$'s are the square roots of the eigenvalues of the positive non-Hermitian operator
$\rho\widetilde\rho$ in decreasing order. We have also introduced the Pauli operator $\sigma_y$ and 
$\widetilde\rho=\sigma_y^{\otimes 2}\rho^\ast\sigma_y^{\otimes 2}$ where $\rho^\ast$ is obtained from $\rho$ after complex conjugation  in the computational basis.
For a pure state, the concurrence reduces to 
$C(\ket\psi)=|\braket{\psi}{\widetilde\psi}|$ with $\ket{\widetilde \psi}=\sigma_y\otimes\sigma_y\ket\psi^\ast$
 and the complex conjugated vector $\ket\psi^\ast$
in the computational basis. By noting that the Bell states fulfill the relations
$\ket{\widetilde\Psi^\pm}=\pm\ket{\Psi^\pm}$, and
$\ket{\widetilde\Phi^\pm}=\mp\ket{\Phi^\pm}$,
it is not hard to realize that  the concurrence of the initial state
in Eq. \eqref{eq:psiin} is  given by
\begin{equation}
	C(\ket\psi)=
	\left|
	c_-^2-d_-^2
	-c_+^2+d_+^2
	\right|.
\end{equation}
This result will serve as guidance for the concurrence of the atomic mixed states after the interaction with the oscillator.

Having introduced the entanglement measure and its initial form, it is now appropriate to evaluate this quantity for the mixed state of the atoms
after the interaction with the mode. 
At odd quarters of the revival time, the expression for the concurrence can be evaluated in closed form using Eqs.~\eqref{eq:rhoat} and \eqref{eq:concurrence}. 
The calculations are somehow tedious as $\braket{\zeta}{\widetilde\zeta_{1,k}}\neq 0$, however, one can find that the only two
nonzero values of $\lambda_{i}$  are given by $(\sqrt{|c_-^2-d_-^2|^2+2|dr|^2}\pm |c_-^2-d_-^2|)/2$. Therefore,
the concurrence reduces to the simple expression
\begin{equation}
	\label{eq:conct4}
	C(\rho_{\rm at}(\tfrac{k}{4}t_{\rm r}))
		=|\braket{\zeta}{\widetilde\zeta}|
	=	
	\left|
	c_-^2-d_-^2
	\right|.
\end{equation}
This results shows that the atomic entanglement at this point has contribution only from the stationary part of the state vector. This is in line with what is expected
from the basin of attraction \cite{Jarvis2009a} that no longer touches a minimum for nonzero  $c_-$ and $d_-$.

At half revival time, the reduced density matrix of the atoms is given 
by a rank two operator with its constituents fulfilling the property $\braket{\zeta}{\zeta_{2,k}}=\braket{\zeta}{\widetilde\zeta_{2,k}}=0$. This feature
enables a simple calculation of the concurrence that is given by
\begin{align}
		\label{eq:conct2}
	C(\rho_{\rm at}(\tfrac{k}{2}t_{\rm r}))&=
		\left||\braket{\zeta}{\widetilde\zeta}|-|\braket{\zeta_{2,k}}{\widetilde\zeta_{2,k}}|\right|
	\\
	&=
    \left|
	\left|
	c_-^2-d_-^2
	\right|-
	\left|d_+^2-c_+^2\right|
	\right|
	\le C(\ket\psi),
	\nonumber
\end{align}
where we have used
$c_{k,\phi}^2-d_{k,\phi}^2={d_\phi^+}^2-c_+^2$.
The last inequality follows from the reverse
triangle inequality and indicates that the entanglement at odd multiples of one half of the revival time cannot be larger than the initial entanglement.
The result can be interpreted as if there was a sort of  competition between the entanglement in the two components leading to a maximum possible entanglement if either
$c_-=d_-=0$ or if $d_+=c_+=0$.
\begin{figure}[t]
	\centering\includegraphics[width=0.42\textwidth]{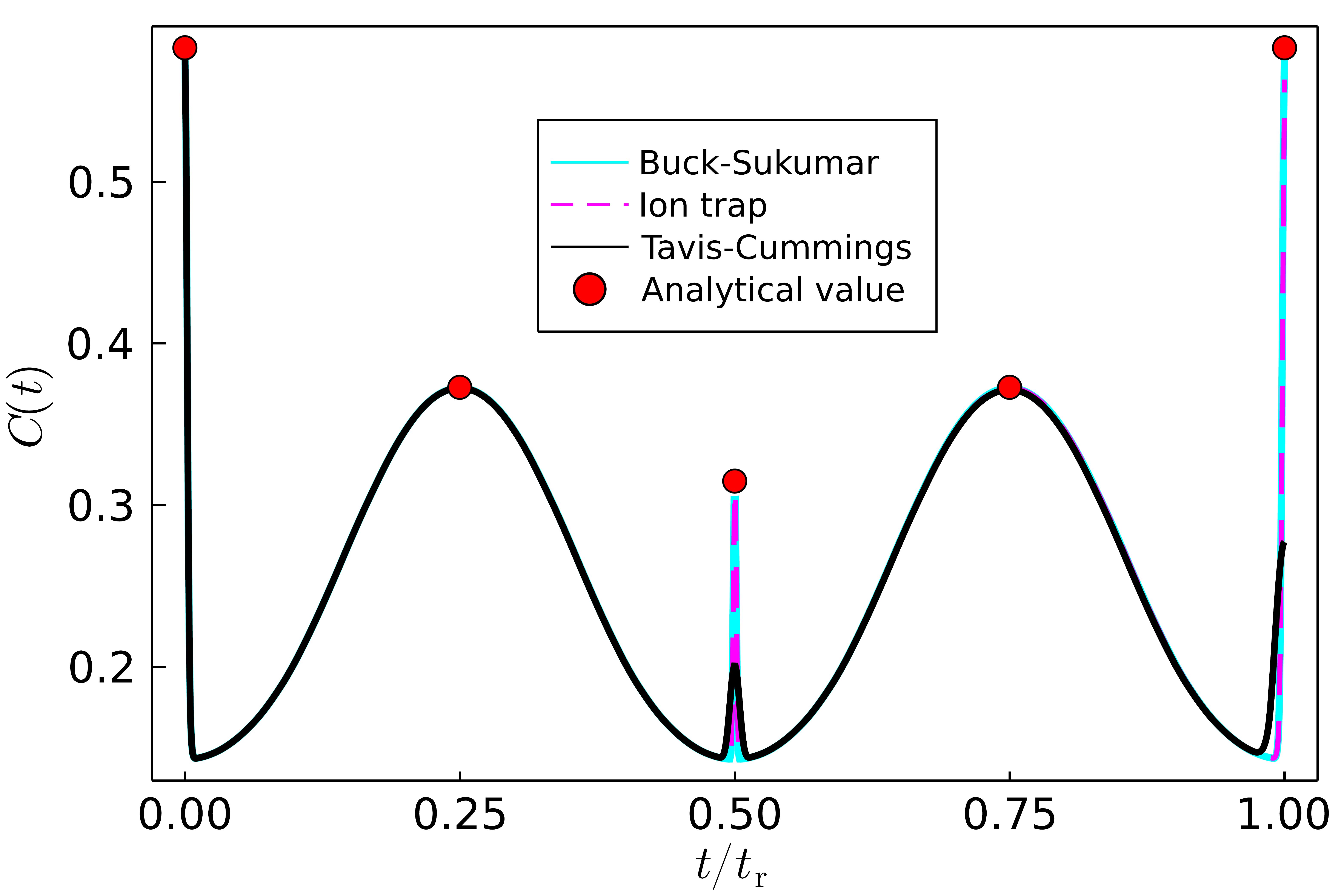}
	\caption{\label{fig:Concurrence}
		Average concurrence as a function of time for the three models as indicated in the legend for $N=2000$. The average was taken from $1000$ random initial
		conditions. The red dots indicate the average analytical value at fixed times given in Eqs. \eqref{eq:conct4} and \eqref{eq:conct4}. The legend indicates the
		curve for each model, that present a substantial overlap. 
	}
\end{figure}
In order to test our analytical prediction, in Fig. \ref{fig:Concurrence} we have plotted the concurrence averaged over $1000$ random initial conditions. Our 
analytical prediction is indicated with a red dot and displays an accurate prediction to the numerical calculation. In all the chosen values of time, we note that
there is a critical point in the behavior of the concurrence; in this case all of them are maximums. This behavior can change, however, depending on the initial
probability amplitudes. 

\subsection{Atoms-oscillator entanglement}
In order to measure the entanglement between both atoms and the oscillator one can use the purity of any of the two density matrices. As we  have already evaluated it for the atomic system in Eq.~\eqref{eq:rhoat},
we will use it to evaluate the purity of the reduced atomic state as
\begin{equation}
	P(t)={\rm Tr}\left\{\rho^2_{\rm at}(t)\right\}.
\end{equation}
Unit value of the purity corresponds to a pure reduced state  and therefore no entanglement between atoms and oscillator. The minimum value of the purity is $1/4$ and corresponds
to a maximally mixed state of the atoms and correspondingly maximum entanglement in the atoms-oscillator bipartition. 

Using Eq.~\eqref{eq:rhoat} it is not difficult to calculate the atomic purity. For odd multiples of one quarter of the revival time the result is given by
\begin{align}
	P\left(\tfrac{kt_{\rm r}}{4}\right)&=p^2+(1-p)(1-|c_-|^2).\\
	p&=|c_-|^2+|d_-|^2.\nonumber
\end{align}
Taking odd multiples of one half of the revival time, one arrives to the following result
\begin{equation}
	P\left(\tfrac{kt_{\rm r}}{2}\right)=p^2+(1-p)^2 \le	P\left(\tfrac{kt_{\rm r}}{4}\right).
\end{equation}
Note that entanglement between the mode and the atoms at odd multiples of $t_{\rm r}/4$ and $t_{\rm r}/2$ depends entirely on the the initial probabilities of
the states $\ket{\Psi^-}$ and $\ket{\Phi^-}$. If non of these states are initially populated, the purity of the atomic reduced density matrix is one and therefore
no entanglement is present in the atoms-oscillator partition at this specific times. Also when $p=1$, the purity takes unit value at these two times. From the previous 
expressions one can find that the minimum value of the purity is $1/2$, attainable for $p^2=1/2$. Therefore it is impossible to maximally entangle
the two atoms with the oscillator. Nevertheless, the amount of achievable degree of entanglement is good enough to generate authentic tripartite entangled states as will
be shown in the next section.

\section{Entangling operations}
\label{sec:EntanglingOperations}
In this section we introduce entangling operations that can be implemented with the aforementioned system and that can be exploited in
quantum information protocols. The results rely on the approximate solution of the state vector in Eq.~\eqref{eq:PsiFinal} for an interaction time
$t_{\rm r}/2$ where the time dependent part takes the simple form in Eq.~\eqref{eq:Psipi2}. As we will be concentrated only in this interaction time,
 it is convenient to introduce the following 
shorthands to be used in this section
\begin{equation}
	\label{eq:shorthands}
	U=e^{-iV t_{\rm r}/2\hbar},\quad \theta=\delta_N t_{\rm r}/2.
\end{equation}
We anticipate that some the resulting two-qubit operations are not unitary, however, they can be of importance in quantum information tasks. For instance, 
it has been shown that one of them can replace the CNOT gate in a recurrence entanglement purification protocol \cite{Bernad2016a,Torres2016a}. This does not impose a major loss
in the protocol, as recurrence purification protocols are probabilistic in nature as the implementation of unitary gates is followed by measurement in the computational basis where half of the results have to be discarded. Here, we will introduce a new entangling operation that can also assist in such a protocol.

\label{QuantumOperations}
\subsection{Two-qubit operations}
Let us first consider a scheme to implement quantum operations based on Bell state projectors. By
inspecting  Eqs.~\eqref{eq:PsiFinal} and \eqref{eq:Psipi2} one can rewrite the approximate solution to the state vector at time $t_{\rm r}/2$ in the following convenient form 
\begin{align}
			\label{eq:Psihalfrev}
	U\ket\psi\ket\alpha\simeq&\left(c_-\ket{\Psi^-} +d_-\ket{\Phi^-}\right)\ket{\alpha}\nonumber\\
+&\left(c_+\ket{\Psi_\theta} +d_+\ket{\Phi_\theta}\right)\ket{-\alpha},
\end{align}
with the orthogonal and maximally entangled  states 
\begin{align}
	\label{eq:Phi1Psi1}
	\ket{\Psi_\theta}&=\cos\theta\ket{\Psi^+}+i\sin\theta \ket{\Phi^+},\nonumber\\
	\ket{\Phi_\theta}&=-i\sin\theta\ket{\Psi^+}-\cos\theta \ket{\Phi^+}.
\end{align}
The coefficients $c_1$ and $d_1$ in Eq. \eqref{eq:ckdk} respectively represent the initial probability amplitudes of these two states. 
We note that a measurement of  the oscillator state $\ket{\alpha}$ or $\ket{-\alpha}$ postselects the atoms in one of two orthogonal states. This would correspond
to two different two-qubit operations. However, projecting on $\ket{-\alpha}$ postselects the atoms in a state that depends on the parameters of the
system and not merely on the initial atomic state. This can be overcome by initially applying a quantum gate that transforms the symmetric Bell states,
while leaving the antisymmetric ones invariant. For this purpose,  we introduce the following unitary gate
\begin{equation}
	G_\theta=\ketbra{\Psi_\theta}{\Psi^+}-\ketbra{\Phi_\theta}{\Phi^+}+\ketbra{\Psi^-}{\Psi^-}+\ketbra{\Phi^-}{\Phi^-}.
\end{equation}
The minus sign in the second element is crucial, as
in this way the required quantum gate is separable and can be expressed in terms of separable (single atom)  gates as
\begin{equation}
	G_\theta=g_\theta\otimes g_\theta,\quad
	g_\theta=
	\cos \tfrac{\theta}{2}\mathbb{I}+i\sin\tfrac{\theta}{2} \sigma_x.
\end{equation}
Using this gate before the interaction, one can obtain the  state at half the revival time given by
\begin{align}
	\label{eq:PsiGTr2}
	U\,G_\theta\,\ket{\psi}\ket{\alpha}\simeq&\left(c_-\ket{\Psi^-} +d_-\ket{\Phi^-}\right)\ket{\alpha}
	\nonumber\\
	+&\left(c_+\ket{\Psi^+} -d_+\ket{\Phi^+}\right)\ket{-\alpha}.
\end{align}
With this result, measuring the state of the oscillator in $\ket{\alpha}$ or $\ket{-\alpha}$ would respectively correspond to the following quantum operations
\begin{align}
	\label{eq:ML}
	M&=\ketbra{\Psi^-}{\Psi^-}+\ketbra{\Phi^-}{\Phi^-}\\
	L&=\ketbra{\Psi^+}{\Psi^+}-\ketbra{\Phi^+}{\Phi^+}.
\end{align}
These Hermitian operators can be regarded as the measurement operators of a positive operator valued measure (POVM) \cite{Nielsen00}. The operators $M$ and $L$  fulfill 
$M^2+L^2=\mathbb{I}$. Furthermore, the sum
of the two of them $M+L$ is a unitary operator. The gate $M$ has already been used in place of the usual CNOT gate in purification protocols \cite{Bernad2016a}. 
The operation $L$ can also be considered in a setting of this type in order to improve the efficiency of these purification protocols. 
A manuscript with these results is also in preparation by two of the authors. 

It is important to comment that for the measurement of the photonic field a projection onto coherent states is not strictly necessary. A projection onto position eigenstates or weighted sum of position eigenstates close to the coherent state would lead to the same atomic postselection. This
 can be implemented using a balanced homdyne measurement as explained in \cite{Torres2014}. 
 In ion traps, one would
require to measure one mode of oscillation of the ions such as the center of mass motion. A drawback in this case is
that a position measurement of an ion destroys its internal
state. In order  overcome this problem,
one could include ancillary ions in the chain that should
not be in contact with the laser generating the interaction
with the center of mass motion. 
At the end of the interaction, the ancillary atoms could be individually addressed \cite{Leibfried2003a,Casabone2013,Haeffner2008} with other lasers in order to measure their position through their fluorescence. Provided that all other modes are cooled down to the ground state, this would correspond to a detection of the center of mass mode. 
As for the single qubit gates $g_\theta$, these 
can, in principle, be implemented by driving the atomic transition with laser pulses and properly
controlling their duration  \cite{Raimond2001,Meschede2006}. In the case of trapped ions, the carrier resonance has to be chosen in order to
avoid excitation of the mechanical mode \cite{Leibfried2003a}.


\subsection{Three-qubit operations}
As we have seen in Sec. \ref{Entanglement}, the dynamics of this model is able to generate simultaneous entanglement between the two atoms and also between these two and  the oscillator.
For this reason, it is natural to expect the possibility of tripartite entanglement in the system. It is known that there are two inequivalent types of tripartite entangled states of three qubits \cite{Dur2000} that can be represented by the following states
\begin{align}
	\label{eq:GHZW}
    \ket{\rm GHZ}&=(\ket{000}+\ket{111})/\sqrt 2,\\
	\ket{W}&=(\ket{001}+\ket{010}+\ket{100})/\sqrt 3.
\end{align}
These two state, $W$-state and the GHZ state, posses entanglement among any bipartition of the three qubits. We will show that it is possible to generate
both of theses paradigmatic tripartite entangled states from an initial separable state and using as entangling gate the evolution operator $U$ in Eq.
\eqref{eq:shorthands}.

Let us first consider the generation of GHZ states, as it follows directly from the solution of the state vector.
Examining Eq. \eqref{eq:PsiGTr2}, it is possible to note that in order to generate a GHZ state, it suffices to initialize both atoms in the ground state, 
where $d_-=d_+=1/\sqrt2$ and $c_-=c_+=0$, and apply the separable atomic  gate $G_\theta$ followed by the evolution
operator $U$. With these conditions one can obtain a GHZ state in the following way
\begin{align}
	\label{eq:GHZ}
	U\,G_\theta\ket{\rm gg}\ket\alpha&=\frac{\ket{\rm gg}\ket{\alpha,-}+\ket{\rm ee}\ket{\alpha,+}}{\sqrt 2},\\
	\ket{\alpha,\pm}&=-(\ket{-\alpha}\pm\ket\alpha)/\sqrt 2.\nonumber
\end{align}
We have introduced the symmetric and antisymmetric Schrödinger cat states $\ket{\alpha,\pm}$ that can be considered as the two states of a two level system. This is
plausible provided that $\alpha$ is large enough in order to have $\braket{\alpha,+}{\alpha,-}\simeq 0$.

The generation of a $W$-state from a separable state is more involved, but also not difficult to achieve
using the unitary evolution. For this case, we will first introduce an effective form of the evolution operator $U$ where its action as
a three-qubit gate is more evident. In this case, we will use for the third qubit the nearly orthogonal states $\ket{\pm\alpha}$. As
the solution of the state vector in Eq. \eqref{eq:Psihalfrev} at the specific time $t_{\rm r}/2$ involves only these two coherent states, one can identify that
the evolution operator connects the coherent state $\ket{\alpha}$ with $\ket{\pm\alpha}$. The same is true for $\ket{-\alpha}$ that is only connected to $\ket{\pm\alpha}$, as
one can note by evaluating the action of the evolution operator on $\ket\psi\ket{-\alpha}$ using the interaction picture as defined in  Eq. \eqref{eq:intpic}. 
This means that $U$ is closed in the  subspace spanned by these two coherent states that can be regarded as states of a third qubit. 
With this in mind and by analyzing Eq. \eqref{eq:Psihalfrev}, it is possible to find the following form of the evolution operator
\begin{align}
	U&\simeq M\otimes \mathbb{I}_a+K\otimes\ketbra{-\alpha}{\alpha}+K^\ast\otimes\ketbra{\alpha}{-\alpha},\\
	K&=\ketbra{\Psi_\theta}{\Psi^+}+\ketbra{\Phi_\theta}{\Phi^+}.\nonumber
\end{align}
This form is only valid for an interaction time $t_{\rm r}/2$ and initial coherent states $\ket{\pm \alpha}$. 
In the previous expression we have used $\mathbb{I}_a$ as the identity operator in the oscillator space.
Although not evident at first glance, the operator
$K$ is Hermitian and fulfills the relation $K^2=L^2$. 
An important feature of this effective 
evolution operator is its evident three-qubit gate character that is suitable to analyze three qubits states. 

In order to generate a $W$-state, we choose the specific value  $\theta=\delta_N t_{\rm r}/2=\pi/4$ of the angle in $\ket{\Psi_\theta}$ and $\ket{\Phi_\theta}$
given in Eqs. \eqref{eq:shorthands} and \eqref{eq:Phi1Psi1}.
 This restricts the value
of the mean number $N$ and it is not possible to achieve in every model. For instance, in the Buck-Sukumar model $\delta_N$ and the revival time are constant. In the ion-trap model, however,
$N$ can be chosen according to Eq. \eqref{eq:ITvalues} in order to achieve $\theta=\pi \delta_N /|\omega_N'| =\pi/4$. Note that for large enough value of the mean number of quanta
$N$, its value can slightly differ from the optimal one in Eq. \eqref{eq:eta}, as the coherent state amplitude are narrowly centered in the linear region of $\omega_n$ (see
Fig. \ref{fig:WnPn} for $N=2000$).
With this in mind, and using the following initial separable condition
\begin{equation}
	\ket{\psi_1}=\ket{\rm g}
	\frac{\ket{\rm g}+i\ket{\rm e}}{\sqrt2}
	\frac{\ket\alpha+\sqrt{2}\ket{-\alpha}}{\sqrt3},
\end{equation}
it is not hard to realize that applying the evolution operator results in
\begin{align}
	U\ket{\psi_1}&=\frac{1}{\sqrt{3}}\ket{\psi_2}\ket\alpha+\sqrt{\frac{2}{3}}\ket{\Psi^+}\ket{-\alpha},\\
	\ket{\psi_2}&=(\ket{\Psi^-}+i\ket{\Phi^-}-i\sqrt2\ket{\Phi^+})/2.\nonumber
\end{align} 
We have introduced  the separable state $\ket{\psi_2}$ that is orthogonal to $\ket{\Psi^+}$. In order to bring this state to a more obvious form of a $W$ state, the next task is to find a separable unitary gate that fulfills $T\ket{\psi_2}=\ket{\rm ee}$ and $T\ket{\Psi^+}=\ket{\Psi^+}$. The problem can be solved with the aid of the following separable gate
\begin{equation}
	T=\gamma g_{\pi/4}^\dagger\otimes \gamma g_{\pi/4}, \quad \gamma = i\ketbra{\rm e}{\rm e}+\ketbra{\rm g}{\rm g}.
\end{equation}
In this way, using the gate $T$ one can immediately find that starting from a separable state $\ket{\psi_1}$, one can obtain
\begin{equation}
	T\,U\ket{\psi_1}=\frac{
	\ket{\rm ee}\ket\alpha+
		\ket{\rm ge}\ket{-\alpha}+
			\ket{\rm eg}\ket{-\alpha}
}{\sqrt 3},
\end{equation} 
which is a more evident form of a $W$-state as introduced in Eq. \eqref{eq:GHZW}. 

We have shown the potential to generate authentic tripartite entangled states using the unitary evolution of the model as entangling operation. Furthermore, it is also
relevant to note that for the generation of the $W$-state, a state similar to a Schr\"odinger cat is needed. Although this might be considered as a drawback, it is evident from
the generated GHZ state in 
Eq. \eqref{eq:GHZ} that this model also offers a direct form of generating Schr\"odinger cat states by measuring the atoms in the computational basis.

\subsection{Bell measurement}
\label{BellMeasurement}
\begin{figure}[t]
	\centering\includegraphics[width=0.47\textwidth]{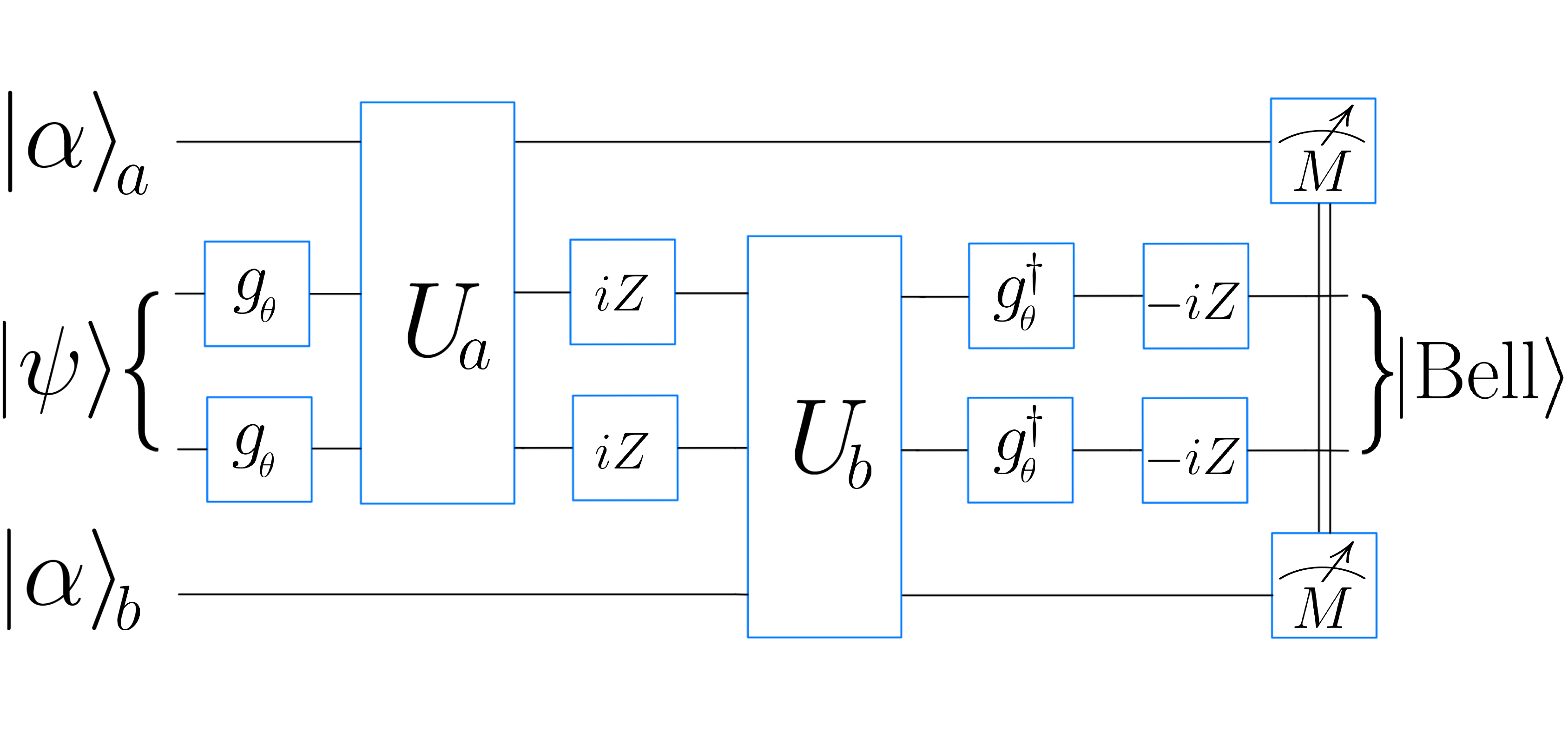}
	\caption{\label{fig:Circuit}
	Quantum circuit representation of the Bell measurement protocol with coherent states 
	$\ket{\pm\alpha}_{a,b}$ used as auxiliary qubits and
	with $e^{iS_z\pi/2}=iZ\otimes iZ$. Here we use the notation $Z=\ketbra{\rm e}{\rm e}-\ketbra{\rm g}{\rm g}$ as the Pauli-$Z$ quantum gate commonly used in quantum computing. At the final stage of the circuit, modes $a$ and $b$ are measured leading
	to four possible outcomes that  postselect the atoms in one
	of the four Bell states as indicated by the state in Eq. \eqref{eq:BellM}.
	}
\end{figure}

Let us briefly sketch  a procedure to implement a Bell measurement by  taking advantage of the interaction dynamics of the
general model introduced above. It is not hard to conceive such a protocol by inspecting the solution for the state vector in 
terms of atomic Bell states and coherent states of the oscillator. 
First let us come back to the state in Eq.~\eqref{eq:PsiGTr2} after applying the gate $G$ on the initial state followed
by the atoms-oscillator interaction $U$. This procedure can be considered as an atomic state splitter, in the sense that one of the material components remains invariant
accompanying the oscillator state $\ket\alpha$, while another material component follows another mode state $\ket{-\alpha}$ that is orthogonal to the first one. 
This interesting feature can be further applied in order to separate again the two components into four components in terms of Bell states. 
In order to do so,
one requires to interchange the Bell states $\ket{\Phi^\pm}$ using a rotation as in Eq. \eqref{eq:EZPi2}, followed by an interaction with an additional oscillator with operators $b$ and $b^\dagger$ described by $V_b$ as in Eq. \eqref{eq:V}. With all these considerations, one can come up with the unitary gate
\begin{equation}
	\label{eq:UBell}
	\mathcal{U}=e^{-iS_z\frac{\pi}{2}}G^\dagger_\theta\,
	U_b
	\,
	e^{iS_z\frac{\pi}{2}}\,
    U_a
	\,G_\theta,
\end{equation}
where we have distinguished between different modes using their annihilation operator as subscript. 
Applying this unitary gate to an initial arbitrary atomic state with two coherent states, one obtains the following state vector with four components
\begin{align}
\label{eq:BellM}
	\mathcal{U}\ket{\psi}\ket{\alpha}_a\ket{\alpha}_b&\simeq
	d_-\ket{\Phi^-}\ket{\alpha}_a\ket{-\alpha}_b+c_-\ket{\Psi^-}\ket{\alpha}_a\ket\alpha_b
	\nonumber\\
	+d_+\ket{\Psi^+}&\ket{-\alpha}_a\ket\alpha_b  +c_+\ket{\Phi^+}\ket{-\alpha}_a\ket{-\alpha}_b.
\end{align}
In this final state, a different combination of coherent states is accompanied by a specific Bell state multiplied by its initial probability amplitude. Therefore, by 
discriminating the four coherent states in the two oscillators, one is able to postselect the atomic state in one of the four Bell states. 
For instance, measuring oscillator states close
to $\ket\alpha\ket{-\alpha}$ corresponds to a projection onto $\ket{\Phi^-}$, as this would happen
with probability $|d_-|^2$. Analogue procedures apply to all four Bell states.
The process can be visualized using the useful circuit representation shown in Fig. \ref{fig:Circuit}.
It is worth commenting, that 
the discrimination of oscillator states need not be
a projection onto coherent states. It suffices a measurement of the oscillator in a localized state close to a specific coherent state.

\section{Conclusions}

We have presented a theoretical analytical study of a nonlinear  intensity dependent two-atom Tavis-Cummings model.
 The exact solvability of the model has been shown
by identifying two constant of motion. By introducing a convenient interaction picture, we have been able to solve the time dependent problem
for initial arbitrary coherent states, using only coherent states that lie in the positive axis in the complex plane. By considering large mean number of quanta
in the oscillator, we have derived an analytical approximate expression given in terms of atomic Bell states and oscillator coherent states that has been
numerically tested using its fidelity with respect to the exact expression. 
As particular cases of this model, we have revised in detail three particular models: the Tavis-Cummings model, the Buck-Sukumar model, and the nonlinear ion-trap model.
We have shown that in the experimentally feasible ion-trap model,  the coherent state approximation can accurately describe the dynamics, when carefully
choosing the mean number of quanta for a given Lamb-Dicke parameter.
The approximate solution of the time-dependent state vector has proven to be very useful in analyzing the dynamical features, and more specifically, the entanglement in the system. The most important result is
that, with the approximate form of the state vector, we have been able to introduce entangling operations for two-qubtis and three qubits in a compact form. 
The results in this work show that the physical implementation of entangling operations relying on nonlinear Tavis-Cummings models can be realized in current ion-trap experiments, opening new avenues for the implementation of basic quantum protocols assisted by multi-phonon states. 

\begin{acknowledgments}
R. G.-R. is grateful to CONACYT for
financial support under a Doctoral Fellowship.
C. A. G.-G. acknowledges funding from the Spanish MICINN through the project MAT2017-88358-C3-1-R, and the program Acciones de Dinamizaci\'on ``Europa Excelencia'' EUR2019-103823. 
J.M.T. acknowledges support from BUAP, project number 100527172-VIEP2021.
The authors would also like to thank Ralf Betzholz for his useful comments on this manuscript.
\end{acknowledgments}

\appendix

\section{Exact form of the evolution operator}
\label{Exact}
For the sake of completeness, in this appendix we briefly present the exact solution of the evolution operator $U(t)=e^{-i Vt/\hbar}$.
The exact expression of each block of $U(t)$ for $n>1$ is given by
\begin{align}
		U^{(n)}(t)=	
	\begin{pmatrix}
		\frac{\Omega_{n-1}^2+\Omega_n^2 \mathcal{C}(t)}{\nu_n^2}& \frac{\Omega_n^2\mathcal{S}(t)}{i\nu_n}&\frac{\Omega_{n-1}\Omega_n( \mathcal{C}(t)-1)}{\nu_n^2}\\
		\frac{\Omega_n^2\mathcal{S}(t)}{i\nu_n}&  \mathcal{C}(t)&\frac{\Omega_{n-1}^2\mathcal{S}(t)}{i\nu_n}\\
		\frac{\Omega_{n-1}\Omega_n( \mathcal{C}(t)-1)}{\nu_n^2}& \frac{\Omega_{n-1}^2\mathcal{S}(t)}{i\nu_n}&\frac{\Omega_{n}^2+\Omega_{n-1}^2 \mathcal{C}(t)}{\nu_n^2}
	\end{pmatrix}
	\nonumber
\end{align}
with the shorthands  $\mathcal{C}(t)=\cos\nu_n t$ and $\mathcal{S}(t)=\sin\nu_n t$, and the exact form of the eigenfrequencies $\nu_n=\sqrt{\Omega_n^2+\Omega_{n-1}^2}$.
For $n=0$ the blocks of the interaction operator $V$ and of the evolution operator $U(t)$ are two dimensional and given by 
\begin{equation*}
	V^{(0)}= 
	\begin{pmatrix}
		0& \hbar\Omega_0\\\hbar\Omega_0&0
	\end{pmatrix}, \, U^{(0)}(t)=
\begin{pmatrix}
	\cos\Omega_0 t & -i \sin \Omega_0 t\\
	-i\sin\Omega_0 t & \cos\Omega_0 t
\end{pmatrix},
\end{equation*}
with the basis states $\ket{\rm gg}\ket{1}$ and $\ket{\Psi^+}\ket{0}$. For $n=-1$, we have only one state, $\ket{\rm gg}\ket0$, and therefore the blocks are one-dimensional with $V^{(-1)}$=0 and $U^{(-1)}$=1.
Using these blocks, the time-dependent amplitudes in Eq. \eqref{eq:Psit} can be evaluated as $C_n(t)=U^{(n)}C_n(0)$ with the column vector $C_n(t)$ containing
the coefficients $C_{n,l}(t)$, $l\in\{-1,0,1\}$. The approximate expressions in Eq. \eqref{eq:Cs} can also be obtained directly from the exact expressions using
the approximations explained  in Sec. \ref{ssec:statevector}.

\section{Derivation of the interaction Hamiltonian in the ion-trap setting}
\label{iontrapV}
In this appendix, we give a brief overview of the ion-laser coupling leading to an interaction Hamiltonian in the form of Eq. \eqref{eq:V} between internal levels of the ions and their center of mass motion. 
We will assume that other normal modes are cooled down to their ground state.  With this condition we can follow the derivation in Refs.  \cite{Vogel1995,Leibfried2003a,Haeffner2008}.

The free Hamiltonian describing the internal levels and center of mass motion of two two-level ions inside a harmonic trap potential is given by
\begin{equation}
\label{eq:H0ions}
    H_0=\hbar\nu a^\dagger a +\hbar\omega S_z.
\end{equation}
Here $a$ and $a^\dagger$ represent the creation and annihilation operators of the center of mass motion of the ions, and 
$\nu$ is the trap frequency. The interaction of the two ions with a monochromatic laser field at frequency $\omega_{\rm L}$ is given by
\begin{equation}
V_L=\hbar\Omega\left(S_-+S_+\right)\cos(k x-\omega_{\rm L} t+\varphi),
\end{equation}
where $k$ is the wave number of the laser and $\Omega$ describes the ion-laser coupling strength. Furthermore, $x$ stands for the position operator of center of mass, and $\varphi$ is an arbitrary phase
fixed by the atomic position with respect to the light wave. Taking into account that $\omega_{\rm L}$ is in the optical regime, the interaction can be simplified using the rotating wave approximation as
\begin{equation}
\label{interactionV}
    V_L=\hbar\Omega S_+e^{i\varphi}
    e^{i\eta(a^\dagger +a)}e^{-i\omega_{\rm L} t}+{\rm H.c.},
\end{equation}
where we have introduced the Lamb-Dicke parameter with the aid of the
relation between position coordinate, the wave number and the bosonic operators as
$kx=\eta(a^\dagger +a)$.
Using the relation 
$e^{i\eta(a^\dagger +a)}=e^{-\eta^2/2}e^{i\eta a^\dagger}e^{i\eta a}$, and the series expansion of the exponential, it is possible to express Eq.\eqref{interactionV} 
in the interaction picture with respect to $H_0$ in \eqref{eq:H0ions} as
\begin{equation*}
    V_{L,I}=\hbar\Omega S_+e^{i\varphi} e^{-\frac{\eta^2}{2}}
    \sum_{l,m=0}^\infty
    \frac{(i\eta)^{l+m}}{l! m!}a^{\dagger\, l}a^m 
    e^{i\Delta_{l,m}t}+{\rm H.c.}
\end{equation*}
with $\Delta_{l,m}=\omega-\omega_{\rm L}-(m-l)\nu$. By choosing
$\varphi=\pi/2$ and tuning the laser frequency to the first red-sideband with respect to the atomic 
transition, i.e., $\omega_{\rm L}=\omega-\nu$, one is able to obtain the following interaction Hamiltonian
\begin{equation}
    V=\hbar\Omega S_+ a \, e^{-\frac{\eta^2}{2}}\eta
    \sum_{l=0}^\infty
    \frac{(-\eta^2)^{l}}{l!(l+1)!}a^{\dagger\, l}a^l +{\rm H.c.}
\end{equation}
which is in the form of Eq. \eqref{eq:V} with $f(a^\dagger a)$
 in Eq. \eqref{nTC:Hamiltonian}.



\end{document}